\documentclass[12pt,draftclsnofoot, perreview, onecolumn]{IEEEtran}
\usepackage{graphicx}
\usepackage{amsmath,amssymb}
\usepackage{mathrsfs}
\usepackage{cases}
\usepackage{color}
\usepackage{amsfonts}
\usepackage{indentfirst}
\usepackage{cite}
\usepackage{caption}
\usepackage{algorithm}
\usepackage{algpseudocode}
\usepackage{booktabs}
\usepackage{subfigure}
\usepackage{multirow}
\usepackage{amsthm}
\usepackage{diagbox}
\makeatletter
\newtheoremstyle{mythm}{3pt}{3pt}{}{16pt}{\bfseries}{:}{.5em}{}
\theoremstyle{mythm}

\begin{document}
%\begin{spacing}{1.0}
%\begin{sloppypar}
\title{Design of Placement Delivery Arrays for Coded Caching with Small Subpacketizations and Flexible Memory Sizes}
\author{Xianzhang Wu, Minquan Cheng, Congduan Li, {\em Member, IEEE}, and Li Chen, {\em Senior Member, IEEE}% <-this % stops a space
\thanks{Xianzhang Wu and Congduan Li are with School of Electronics and Communication Engineering, Sun Yat-sen University, Shenzhen 518107, China (e-mail: wuxzh7@mail2.sysu.edu.cn, licongd@mail.sysu.edu.cn).}
\thanks{Minquan Cheng is with Guangxi Key Lab of Multi-source Information Mining $\&$ Security, Guangxi Normal University,
Guilin 541004, China (e-mail: chengqinshi@hotmail.com).}% <-this % stops a space
\thanks{Li Chen is with School of Electronics and Information Technology, Sun Yat-sen University, Guangzhou 510006, China (e-mail: chenli55@mail.sysu.edu.cn).}
}
%\markboth{IEEE Transactions on Communications, submitted}
\maketitle
\begin{abstract}
 Coded caching is an emerging technique to reduce the data transmission load during the peak-traffic times. In such a scheme, each file in the data center or library is usually divided into a number of packets to pursue a low broadcasting rate based on the designed placements at each user's cache. However, the implementation complexity of this scheme increases as the number of packets increases. It is crucial to design a scheme with a small subpacketization level, while maintaining a relatively low transmission rate. It is known that the design of caches in users (i.e., the placement phase) and broadcasting (i.e., the delivery phase) can be unified in one matrix, namely the placement delivery array (PDA). This paper proposes a novel PDA construction by selecting proper orthogonal arrays (POAs), which generalizes some known constructions but with a more flexible memory size. Based on the proposed PDA construction, an effective transformation is further proposed to enable a coded caching scheme to have a smaller subpacketization level. Moreover, two new coded caching schemes with the coded placement are considered. It is shown that the proposed schemes yield a lower subpacketization level and transmission rate over some existing schemes.
 \end{abstract}
\IEEEpeerreviewmaketitle
\section{Introduction}\label{introduction}
%\IEEEPARstart{T}{he} fast growing demands of video streaming services can easily cause severe network congestions during Internet peak hours. One possible solution is to exploit the off-peak network resources, such as to cache some of the possibly demanded contents in users' local memories, i.e., caches. This is a natural way to utilize each user's own cache to decrease the network traffic when the cached files are requested. The gain offered by this approach, which is called {\it local gain}, depends on the local cache size. A more effective way of caching, namely the {\it coded caching}, proposed by Maddah-Ali and Niesen in \cite{1}, further reduces the network pressure during the peak hours by strategically designing the contents cached into network users to obtain the {\it global gain}.
\IEEEPARstart{T}{he} dramatically increasing demands of video streaming services generate a great challenge to the central servers for a smooth data transmission, especially during the peak hours. Coded caching system has been proposed as a promising technology to reduce the data transmission load during the peak hours by utilizing ample cache memories that are available at the users. The earliest centralized coded caching scheme was proposed by Maddah-Ali and Niesen \cite{1}, which is called the MN scheme in this paper. In the scheme, the network model consists of
a central server containing $N$ files with equal size, which provides service to $K$ users over an error-free broadcasting channel. Each user is assumed to have a cache memory with a size of $M$ files. A coded caching scheme consists of two phases: the placement phase, which occurs during the off-peak hours, and the delivery phase, which occurs during the peak hours. In the placement phase, the server sends the properly designed contents to the cache of each user without knowledge of the demands. In the delivery phase, each user requests one arbitrary file, and the server broadcasts some coded packets to them so that each user can recover its desired file with the assistance of the contents in its own cache. The worst case minimal broadcasting load over all possible demands is defined as the {\it transmission rate} (or {\it rate}) $R$, i.e., the least number of files that must be communicated so that any demands can be satisfied.
A coded caching scheme is called an $F$-division scheme if each file can be equally divided into $F$ packets, which is called the {\it subpacketization level}.
If the packets are cached directly without coding or mixing in the placement phase, it is called the {\it uncoded placement}. Otherwise, it is called the {\it coded placement}.

%\subsection{The Existing Work}
 It has been shown that the MN scheme achieves the optimal rate under uncoded placement when $K\leq N$ \cite{2}, and it is generally order optimal
within a factor of 4 \cite{3}.
%Furthermore, If a file is requested multiple times, Yu {\em et al.} \cite{3} showed that the rate of such scheme can be order optimal within a factor of 2 by removing the redundant transmissions in MN scheme.
 However, subpacketization level of the scheme increases exponentially with the number of users $K$, which makes it infeasible in practice. Therefore, in the centralized coded caching scheme, it is important to reduce the subpacketization level, while maintaining a relatively low transmission rate.
%\subsection{Prior Work}

%It is well known that with a fixed number of users $K$ and cache memory ratio $\frac{M}{N}$, there exists a tradeoff between the transmission rate $R$ and the subpacketization level $F$.
There exist some works on reducing the subpacketization level of the MN scheme while increasing the
transmission rate as a tradeoff \cite{20,22,21,16,17,18,26,13,14,50,39,19,29,24,23,31,33}.
For example, Yan {\it et al}. \cite{13} proposed a combinatorial structure called the placement delivery array (PDA), and showed that the MN scheme can be represented by a special PDA. Shangguan {\it et al}. \cite{16} later showed that many previously existing coded caching schemes could also be represented by the PDA. With the introduction of PDA, various coded caching schemes with a lower subpacketzation level than the MN scheme were proposed in \cite{39,19,29,24,23,31,33}. Other combinatorial construction methods to reduce the subpacketization level were realized by the use of projective geometry \cite{21}, Ruzsa-Szemer\'{e}di graphs \cite{17}, hypergraphs \cite{16}, linear block codes \cite{50}, strong edge coloring of bipartite graphs \cite{14}, and combinatorial design theory \cite{20}. Table I summarizes some known deterministic schemes with advantages in either the subpacketization level or the transmission rate. In Table I, $\left[k \atop t\right]_q=\frac{(q^{k}-1)\cdots (q^{k-t+1}-1)}{(q^{t}-1)\cdots (q-1)}$ for any positive integers $k, t$ and a prime power $q$.
%in \cite{18}, Shanmugam {\em et al.} pointed out that all the deterministic $F$-division coded caching schemes can be recast into an $F\times K$ combinatoric structure, which is called the placement delivery array (PDA) \cite{14}. Later, Cheng {\em et al.} \cite{39} showed that given the minimum rate, the MN scheme achieves the minimum supacketization. However, the subpacketization is still too large for practical scenarios. Shangguan {\em et al.} \cite{16} reduced the subpacketization of the MN scheme by increasing the transmission rate based on hypergraphs and partitions. Cheng {\em et al.} \cite{19} expanded the construction of \cite{16} and obtained the coded caching scheme with a flexible memory size. They further improved the subpacketization of \cite{16} with the same number of users, memory ratio and transmission rate based on the orthogonal arrays (OAs) \cite{29}. Zhang {\em et al.} \cite{24} improved the schemes in \cite{14}, considering the scenario where some packets cached in users do not generate multicasting opportunities in the delivery phase. There are some other methods to construct the coded caching schemes, such as concatenating construction \cite{26,31,23}, bounded subsets \cite{32}, and combinations of strong edge colorings \cite{33}, etc. Table I summarizes some known deterministic schemes with advantages in the subpacketization or the transmission rate. Note that in Table I, $\left[k \atop t\right]_q=\frac{(q^{k}-1)\cdots (q^{k-t+1}-1)}{(q^{t}-1)\cdots (q-1)}$ for any positive integers $k, t$ and prime power $q$.
\begin{center}
\includegraphics[scale=0.2185]{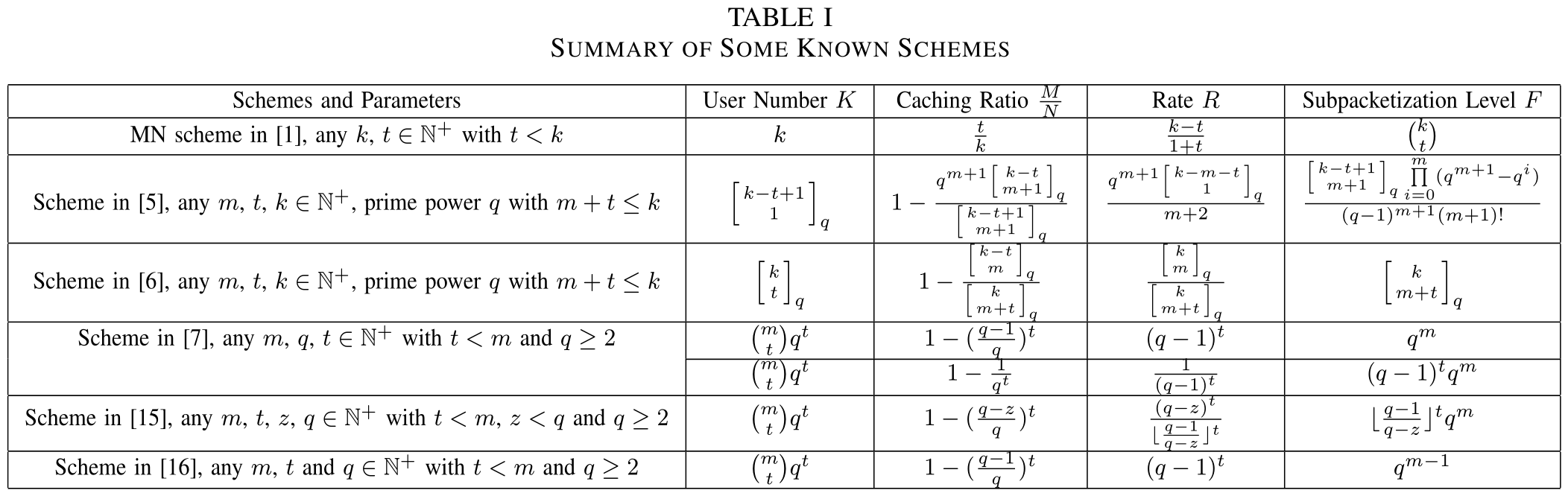}
%\caption{\label{2}Coded caching system.}
 \end{center}

This paper considers the construction of PDA for centralized coded caching scheme, aiming to achieve a low transmission rate with a smaller subpacketization level and a more flexible cache size. Its technical contributions are in three folds.

 %First, by defining the proper orthogonal array (POA), we propose a novel construction of PDAs, based on which, the novel coded caching scheme (as stated in Theorem 1) can reduce the number of packets by a factor of $q$ without sacrificing the transmission
%rate, compared with the existing scheme of \cite{19}, and yield a more flexible memory size, compared with the scheme of \cite{29}, as listed in the first row of Table II. The proposed construction can be regarded as a generalization of \cite{29}, but it highly depends on smartly selecting POAs, which also differs the construction of \cite{19}.
%Unlike the construction of \cite{19}, i.e., the row indices of its PDA are obtained by just reusing the row indices of the PDA of \cite{16}, our construction is done by carefully selecting the proper OAs.
First of all, by defining the proper orthogonal array (POA), we propose a novel construction of PDAs, based on which, the corresponding coded caching scheme (as stated in {\it Theorem 1}) can reduce the number of packets by a factor of $q$ without sacrificing the transmission
rate, compared with the existing scheme of \cite{19}. It also yields a more flexible memory size over the scheme of \cite{29}. Table II presents the advanced features of the proposed PDA schemes, where the conclusion of {\it Theorem 1} can be seen. The proposed construction can be seen as a generalization of the work of \cite{29}, but it requires a delicate selection of the POAs. %which also differs the construction of \cite{19}.

 Secondly, elaborating the use of POAs, we present an effective transformation which leads to another coded caching scheme (as stated in {\it Theorem 2} and also summarized in Table II). By appropriately designing the baseline arrays to satisfy the PDA constraints, we show that when the cache memory ratio is greater than $\frac{1}{2}$, the proposed scheme can work for a larger number of users and a smaller subpacketization level compared with the scheme in {\it Theorem 1}.

 Finally, by observing that in the proposed PDAs, some packets have no multicasting opportunities in the delivery phase, we modify the uncoded placement into coded placement. Consequently, two coded caching schemes (as stated in {\it Theorems 3} and {\it 4}, and summarized in Table II) with smaller subpacketization levels and transmission rates are further proposed.
%This paper follows the steps of \cite{29}. We find that the special OAs, i.e., the proper OAs, can be utilized to design a PDA scheme with a flexible memory size. It will be shown that designing such coded caching schemes highly depends on smartly selecting the row indices of array under the PDA constraints. From this point of view, we present two classes of PDA schemes by carefully selecting the proper OAs, one from expanding the construction of \cite{29}, and the other from some $``\ast"$s of the proposed PDAs without generating multicasting opportunities in the delivery phase. By analyzing our schemes, it is further shown that the new PDA schemes can significantly reduce the subpacketization level over the existing scheme in \cite{19}. The new PDA schemes can also yield a more flexible memory size, providing a wider range of applications.
%This paper proposes two new centralized coded caching PDA schemes by using the proper OAs. The designs can lead to a lower subpacketization level and a more flexible memory size. Since the problem of designing such coded caching schemes can be converted to finding the appropriate row indices of array under the PDA constraints, we utilize the proper OAs to generate the row indices of these arrays. Our analysis will show that the new PDA schemes can significantly reduce the subpacketization level over the existing scheme in \cite{19}. The new PDA scheme can also yield a more flexible memory size, providing a wider range of applications.
\begin{center}
\includegraphics[scale=0.2185]{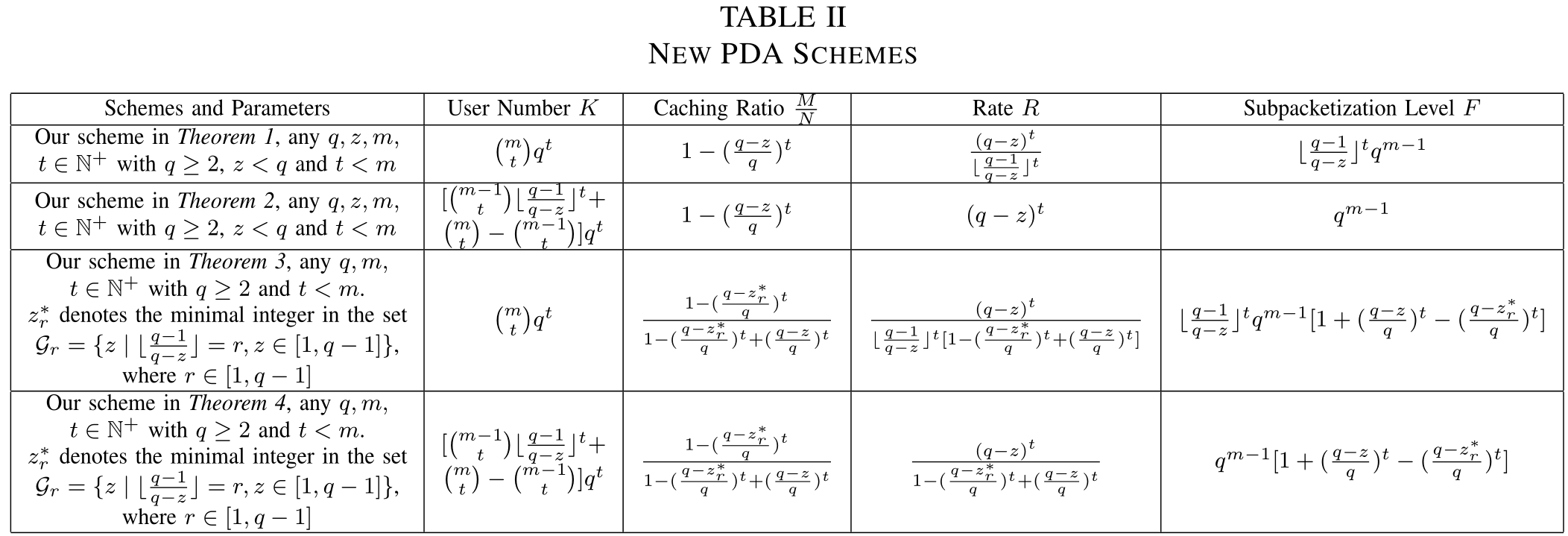}
%\caption{\label{2}Coded caching system.}
 \end{center}

The rest of the paper is organized as follows. Section II formulates
the coded caching problem and reviews some related results.
The new PDA schemes are presented in Section III. Performance analyses of the proposed PDA schemes are given in Section IV. Finally, Section V concludes the paper.
\section{System Model and Related Results}
%In this section, we first state the centralized coded caching system and its relationship to PDA. Then two previously known PDA-based coded caching %schemes via OAs are introduced.
This section briefly reviews the centralized coded caching system and some existing PDA constructions via OAs. Some key notations are also introduced as follows.

\textbf{Notations}: Bolded capital letters denote arrays, bolded lower-case letters denote vectors, and curlicue letters denote sets. Symbol $\oplus$ represents exclusive-or (XOR) operation. Let $\mathbb{N}^{+}$ denote the set of positive integer. The sets of consecutive integers are denoted as $[x, y]=\{x, x+1,\cdots, y\}$.
We use ${[0, m-1]\choose t}$ to represent the collection of all subsets of $[0, m-1]$ with size $t$, i.e., ${[0, m-1]\choose t}=\{\mathcal{S}\mid\mathcal{S}\subseteq[0, m-1], |\mathcal{S}|=t\}$.
%For an $m$-length vector $\textbf{f}$ and a set $\mathcal{S}\subseteq[0, m-1]$, let $\textbf{f}|_{\mathcal{S}}$ denote a vector obtained by deleting the coordinates $j\in[0, m-1]\setminus \mathcal{S}$.
Given an $l\times m$ matrix \textbf{F} and a subset $\mathcal{S}\subseteq[0, m-1]$, let $\textbf{F}|_{\mathcal{S}}$ denote a submatrix of \textbf{F}, which is obtained by taking all the columns indexed by $j\in\mathcal{S}$. Further let $(\textbf{A}_0;\textbf{A}_1;\cdots;\textbf{A}_n)$ denote an array obtained by arranging arrays (or row vectors) $\textbf{A}_0,\textbf{A}_1,\cdots,\textbf{A}_n$ from top to bottom, e.g., $(\textbf{A}_0;\textbf{A}_1)=\left(
  \begin{array}{c}
  \textbf{A}_0\\
    \textbf{A}_1 \\
  \end{array}
\right)$.
Finally, all the vectors in examples are written as strings,
e.g., (1, 0, 1, 0) is written as (1010).
%Note that arrays, vectors and sets are represented by bolded capital letters, bolded lower-case letters and curlicue letters, respectively. The symbol $\oplus$ represents bit-wise XOR. Let $\mathbb{N}^{+}$ denote the positive integer set, while $[x, y]=\{x, x+1,\cdots, y\}$ denotes the interval of integers between two integers $x$ and $y$ $(x<y)$ with different boundary elements. Given $m, t\in\mathbb{N}^{+}$ with $t<m$, we denote ${[0, m-1]\choose t}=\{\mathcal{S}\mid\mathcal{S}\subseteq[0, m-1], |\mathcal{S}|=t\}$. For an $m$-length vector $\textbf{f}$ and a set $\mathcal{S}\subseteq$$[0, m-1]$, $\textbf{f}|_{\mathcal{S}}$ is a vector obtained by deleting the coordinates $j\in[0, m-1]\setminus \mathcal{S}$. Moreover, vectors are written as strings, e.g., (1, 0, 1, 0) is written as (1010).
\subsection{Centralized Coded Caching System}
In a centralized coded caching system, a server
containing $N$ files with equal size is connected to $K$ users through an error-free shared link, as shown in Fig.1. Each
user has a cache of size $M$ files, where $M < N$. The $N$ files and $K$ users are denoted by
$\mathcal{W} =\{W_0, W_1, \cdots, W_{N-1}\}$ and $\mathcal{K}=[0, K-1]$, respectively. An $F$-division $(K, M, N)$ coded caching scheme consists of two phases as follows.
\begin{figure}[H]
\centering
\includegraphics[scale=0.21]{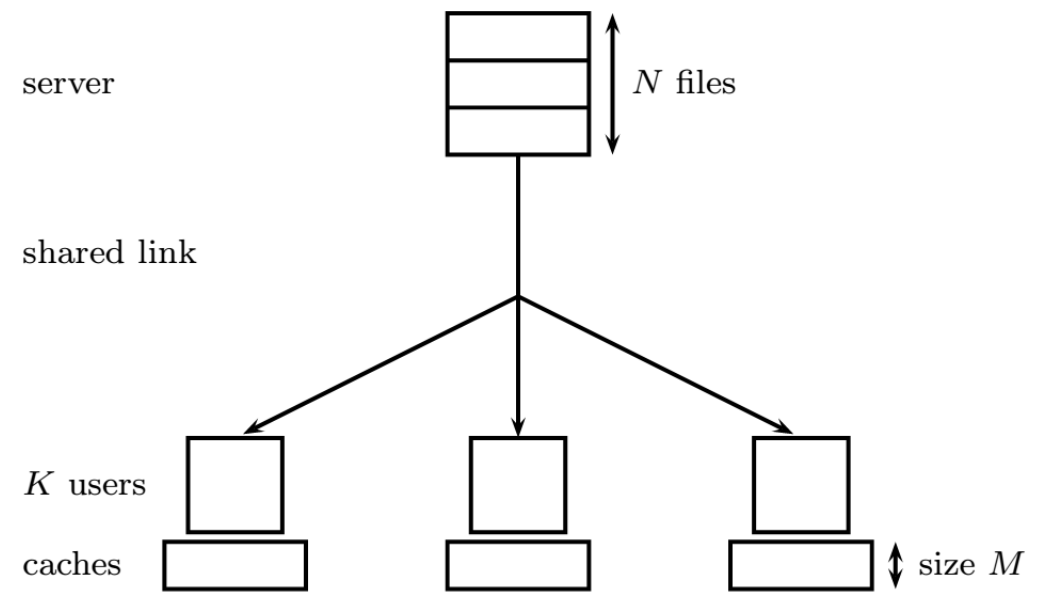}
\caption{\label{2}Coded caching system.}
 \end{figure}

$\bullet$ \textbf{Placement Phase}: Each file is divided into $F$ equal packets, i.e.,
%\footnote{Memory sharing technique may lead to non equally divided packets \cite{1}, in this paper, we will not discuss this case.}
$W_n = \{W_{n,j}| j \in[0, F-1]\}$, $n\in[0,N-1]$. Each user can access the files set $\mathcal{W}$. Let $\mathcal{Z}_k$ denote the packet subset of $\mathcal{W}$ cached by user $k$, where $k\in \mathcal{K}$. Note that the size of $\mathcal{Z}_k$ cannot be greater than each user's cache memory size $M$, i.e., $|\mathcal{Z}_k|\leq M$.

$\bullet$ \textbf{Delivery Phase}: Each user requests one arbitrary file in $\mathcal{W}$. The request vector is denoted by
$\textbf{d}=(d_0, d_1, \cdots, d_{K-1})$, i.e., user $k$ requests file $W_{d_k}$, where $k\in \mathcal{K}$ and $d_k\in[0, N-1]$. Once the
server receives the request vector $\textbf{d}$, it broadcasts at most $RF$ packets such that each user can recover its requested file together with the contents in its own cache.

The above two phases can be described by the PDA that is defined below.

\emph{\textbf{Definition 1}} \cite{13}. Given $K, F, Z, S\in\mathbb{N}^{+}$, an $F\times K$ array $\textbf{P}=(p_{i,j})$ ($i\in[0, F-1]$ and $j\in[0, K-1]$), which consists of a
special symbol $``\ast"$  and $S$ nonnegative integers $0, 1, \cdots, S-1$, is called a $(K, F, Z, S)$ PDA if it satisfies the following conditions:

C1. Symbol $``\ast"$ appears exactly $Z$ times in each column;

C2. Each integer of $[0,S-1]$ appears at least once in the array;

C3. For any two distinct entries $p_{i_1,j_1}$ and $p_{i_2,j_2}$ , $p_{i_1,j_1} = p_{i_2,j_2} = s$ is an integer only if

(a). $i_1\neq i_2, j_1 \neq j_2$, i.e., they lie in distinct rows and distinct columns; and

(b). $p_{i_1,j_2} = p_{i_2,j_1} = \ast$, i.e., the corresponding $2\times2$ subarray formed by rows $i_1, i_2$ and columns $j_1, j_2$ must be in one of
the following forms
\begin{equation*}
\left(
  \begin{array}{cc}
    s & \ast \\
    \ast & s \\
  \end{array}
\right) \text{,}\; \left(
  \begin{array}{cc}
    \ast & s \\
    s & \ast \\
  \end{array}
\right).
\end{equation*}

For example, the following array is a (4, 4, 2, 4) PDA.
\begin{equation}
\textbf{P}=\left(
  \begin{array}{cccc}
    0 & \ast & 2& \ast\\
    \ast & 0& \ast& 2 \\
    \ast & 1& \ast& 3 \\
   1 &\ast& 3&\ast \\
  \end{array}
  \right).
\end{equation}
\begin{algorithm}[!h]
\caption{Coded Caching Scheme Based on PDA in \cite{13}}
%\begin{algorithmic}[1]
\textbf{1:} \textbf{Procedure} Placement (\textbf{P}, $\mathcal{W}$)\\
\textbf{2:} $\;\;\;\;$Split each file $W_n\in \mathcal{W}$ into $F$ packets as $W_n=\{W_{n,j}\mid j\in[0, F-1]\}$.\\
\textbf{3:} $\;\;\;\;$\textbf{For} {$k\in\mathcal{K}$} \textbf{do}\\
\textbf{4:} $\;\;\;\;\;\;$ $\mathcal{Z}_k\leftarrow\{W_{n,j}\mid p_{j,k}=\ast, \forall n\in[0, N-1]\}$;\\
%\STATE \textbf{end procedure}
\textbf{5:} \textbf{Procedure} Delivery$\;(\textbf{P}, \mathcal{W}, \textbf{d})$\\
\textbf{6:} $\;\;\;\;$\textbf{For} {$s=0, 1, \cdots, S-1$} \textbf{do}\\
\textbf{7:} $\;\;\;\;\;\;$ Server sends $\oplus_{{p_{j, k=s, j\in[0, F-1], k\in[0,K-1]}}}W_{d_k, j}$.
%\ENDFOR
%\STATE \textbf{end procedure}
\label{code:recentEnd}
%\end{algorithmic}
\end{algorithm}

 Algorithm 1 has been introduced to realize the PDA based coded caching schemes in \cite{13}. Given a $(K, F, Z, S)$ PDA \textbf{P} with columns representing the user indices and rows representing the packet indices, if
$p_{j,k} = \ast$, user $k$ has cached the $j$th packet of all the files. Condition C1 of {\it Definition 1} implies that all the users have the same memory size and the memory ratio is $\frac{M}{N}=\frac{Z}{F}$. If $p_{j,k} = s$ where $s\in [0, S-1]$, the $j$th packet of all the files is not cached by user $k$. The XOR of the requested packets indicated by $s$ will be broadcast by the server at time slot $s$. Condition C3 of {\it Definition 1} guarantees that user $k$ can get its required packet, since it has cached all the other packets in the multicast message except the requested one. Finally, Condition
C2 of {\it Definition 1} implies that the number of packets transmitted by the server is exactly $S$ and the transmission rate is $R =\frac{S}{F}$.
Furthermore, the coding gain in each time slot $s\in[0, S-1]$, denoted by $g_s$, equals to the occurrence number of integer $s$ in \textbf{P}. This is because the coded packet broadcast at time slot $s$ is useful for $g_s$ users.
Based on Algorithm 1, the following lemma can be obtained.

\emph{\textbf{Lemma 1}} \cite{13}. An $F$-division coded caching scheme for a $(K, M, N)$ caching system can be realized by a $(K, F, Z, S)$ PDA with a memory ratio of $\frac{M}{N}=\frac{Z}{F}$ and a transmission rate of $R =\frac{S}{F}$.

Therefore, based on Algorithm 1, the PDA \textbf{P} in (1) can realize a 4-division (4, 2, 4) coded caching scheme as follows.

$\bullet$ \textbf{Placement Phase}: Each file $W_n$ is divided into 4 packets, i.e., $W_n=\{W_{n,0}, W_{n,1}, W_{n,2}, W_{n,3}\}$, where $n\in[0, 3]$. The contents cached by each user are
\begin{equation*}
\begin{aligned}
&\mathcal{Z}_0=\{W_{n,1}, W_{n,2}\mid n\in[0, 3]\};\mathcal{Z}_1=\{W_{n,0}, W_{n,3}\mid n\in[0, 3]\};\\
&\mathcal{Z}_2=\{W_{n,1}, W_{n,2}\mid n\in[0, 3]\};
\mathcal{Z}_3=\{W_{n,0}, W_{n,3}\mid n\in[0, 3]\}.
   \end{aligned}
\end{equation*}

$\bullet$ \textbf{Delivery Phase}: Assume that the request vector is $\textbf{d}=(0, 1, 2, 3)$. The packets sent by the server at
four time slots are listed as follows. Time slot 0: $W_{0,0}\oplus W_{1,1}$; Time slot 1: $W_{0,3}\oplus W_{1,2}$; Time slot 2: $W_{2,0}\oplus W_{3,1}$; Time slot 3: $W_{2,3}\oplus W_{3,2}$. Then each user can recover its requested
file. E.g., user 0 requests the file $W_0=\{W_{0,0}, W_{0,1}, W_{0,2}, W_{0,3}\}$ and has cached
$W_{0,1}$ and $W_{0,2}$. At time slot 0, it can receive $W_{0,0}\oplus W_{1,1}$, then it can recover $W_{0,0}$ since it
has cached $W_{1,1}$. At time slot 1, it can receive $W_{0,3}\oplus W_{1,2}$, then it can recover $W_{0,3}$ since
it has cached $W_{1,2}$. The transmission rate is $R=\frac{4}{4}=1$.
\subsection{Orthogonal Arrays and Proper Orthogonal Arrays}
%We first give the definition of orthogonal array.
\emph{\textbf{Definition 2}} \cite{30}. Given $m, q, t\in\mathbb{N}^{+}$ with $q\geq2$ and $t\leq m$, let \textbf{F}
%\begin{equation*}
%\textbf{F}=
% \left(\begin{matrix}
 %    \textbf{f}_0\\
%     \textbf{f}_1\\
 %    \vdots\\
 %    \textbf{f}_{l-1}
%\end{matrix}\right)=\left(\begin{matrix}
%    f_{0,0}        & f_{0,1}        &\cdots     & f_{0,m-1} \cr
%    f_{1,0}        & f_{1,1}        &\cdots     & f_{l,m-1} \cr
%  \vdots    &\vdots   &\cdots     &\vdots\cr
%    f_{l-1,0}        & f_{l-1,1}        &\cdots     & f_{l-1,m-1} \cr
%\end{matrix}\right)
%\end{equation*}
denote an $l\times m$ matrix over $[0, q-1]$. It
is called an orthogonal array with a strength of $t$, denoted by OA$_{\lambda}(l, m, q, t)$, if for each subset $\mathcal{S}\in{[0,m-1]\choose{t}}$ with size $t$, every $t$-length $(t\leq m)$ row vector appears exactly $\lambda = \frac{l}{q^{t}}$ times in $\textbf{F}|_{\mathcal{S}}$.
%\begin{equation*}
%\textbf{F}|_{\mathcal{S}}=
% \left(\begin{matrix}
%    \textbf{f}_0|_{\mathcal{S}}\\
 %   \textbf{f}_1|_{\mathcal{S}}\\
  %  \vdots\\
  %  \textbf{f}_{l-1}|_{\mathcal{S}}
%\end{matrix}\right).
%\end{equation*}

Since $l = \lambda q^{t}$ for any OA$_{\lambda}(l, m, q, t)$, it can be simplified into OA$_{\lambda}(m, q, t)$, where $\lambda$ is the index of the orthogonal array. Note that if $\lambda=1$, it can be omitted. E.g., with $m=3, q=2$, and $t=2$, we can consider the following matrix
\begin{equation}
\textbf{F}=(\textbf{f}_0;\textbf{f}_1;\textbf{f}_2;\textbf{f}_3)=((110);(000);(101);(011)).
\end{equation}
 %\begin{equation}
%\textbf{F}=\left(\begin{matrix}
%     \textbf{f}_0\\
%   \textbf{f}_1\\
% \textbf{f}_2\\
%    \textbf{f}_{3}
%    \end{matrix}\right)=
% \left(\begin{matrix}
%     1& 1& 0\cr
%       0 & 0& 0\cr
%    1 & 0& 1\cr
%   0 & 1& 1\cr
%\end{matrix}\right)
%\end{equation}
For each $\mathcal{S}\in{[0,2]\choose{2}}$, we have\\
$\textbf{F}|_{\{0, 1\}}=((11);(00);(10);(01)), \textbf{F}|_{\{1, 2\}}=((10);(00);(01);(11)), \textbf{F}|_{\{0, 2\}}=((10);(00);(11);(01)).$
%\begin{equation*}
%\textbf{F}|_{\{0, 1\}}=
% \left(\begin{matrix}
%    1& 1\cr
 %   0& 0\cr
 %   1& 0\cr
 %   0& 1\cr
%\end{matrix}\right),
%\textbf{F}|_{\{1, 2\}}=
% \left(\begin{matrix}
  %  1& 0\cr
%    0& 0\cr
%    0& 1\cr
%    1& 1\cr
%\end{matrix}\right),
%\textbf{F}|_{\{0, 2\}}=
% \left(\begin{matrix}
 %   1& 0\cr
 %   0& 0\cr
 %   1& 1\cr
%    0& 1\cr
%\end{matrix}\right).
%\end{equation*}
 It can be seen that \textbf{F} in (2) satisfies {\it Definition 2}. Hence, it is an OA$(3, 2, 2)$.

 Based on the definition of OA, we also need a particular type of OA, that is the proper OA (POA), which will enable us to design the new PDAs.

 \emph{\textbf{Definition 3}}. Given $m, q\in\mathbb{N}^{+}$ with $q\geq2$ and $m\geq 2$, an OA$(m, q, m-1)$ is called a proper OA, denoted by POA$(m,q,m-1)$, if the sum (mod $q$) of each row is a constant.

Since the POAs are crucial to our construction, as will be shown later, we need to show the existence of POAs. In fact, it is true that the POAs always exist for any integers $m$ and $q$, where $m\geq 2$ and $q\geq 2$.

 \emph{\textbf{Lemma 2}}. Let \textbf{F} denote a $q^{m-1}\times m$ matrix over $q$ with the set of all rows given as
\begin{equation}
 \mathcal{F}=\{(f_{0}, f_{1}, \cdots, f_{m-2}, x-\sum\limits_{i=0}^{m-2}f_{i})\mid f_{0}, f_{1}, \cdots, f_{m-2}\in[0, q-1]\},
\end{equation}
where $x\in[0, q-1]$, and $m$ and $q$ are greater than 2, then \textbf{F} is a POA$(m, q, m-1)$.
\begin{proof}
 Given any subset $\mathcal{S}\in{[0,m-1]\choose{m-1}}$, if $\mathcal{S}=[0,m-2]$, it can be seen that every $(m-1)$-length row vector appears exactly once in $\textbf{F}|_{\mathcal{S}}$. Let us consider $\mathcal{S}=[0,m-1]\backslash\{j\}$, where $j\in[0,m-2]$. In order to show that every $(m-1)$-length row vector appears exactly once in $\textbf{F}|_{\mathcal{S}}$ for such $\mathcal{S}$, one needs to show that it is impossible for an $(m-1)$-length row vector to appear more than once in $\textbf{F}|_{\mathcal{S}}$, since the total number of row vectors with length $(m-1)$ is $q^{m-1}$. Suppose that there exists an $(m-1)$-length row vector appearing more than once in $\textbf{F}|_{\mathcal{S}}$. Without loss of generality, we can assume that $\textbf{f}|_{\mathcal{S}}=\textbf{f}'|_{\mathcal{S}}$ and $\textbf{f}\neq \textbf{f}'$, where $\textbf{f}=(f_{0}, f_{1}, \cdots, f_{m-2}, x-\sum_{i=0}^{m-2}f_{i})\in\mathcal{F}$ and $\textbf{f}'=(f'_{0}, f'_{1}, \cdots, f'_{m-2}, x-\sum_{i=0}^{m-2}f'_{i})\in\mathcal{F}$. This implies that $f_{j}=f_{j}'$, i.e., $\textbf{f}=\textbf{f}'$, which contradicts to the hypothesis. Therefore, \textbf{F} is an OA$(m,q,m-1)$. Furthermore, the sum of each row of \textbf{F} equals to $x$. Hence, it is also a POA$(m,q,m-1)$.
\end{proof}
It can be seen that the matrix in (2) is a POA$(3, 2, 2)$ since the sum of each row is 0. It is worth pointing out that an OA$(m, q, m-1)$ is not always a POA$(m, q, m-1)$. E.g., with $m=3, q=3$, and $t=2$, the following matrix \textbf{F} is an OA$(3, 3, 2)$, but it is not a POA$(3, 3, 2)$.
\begin{equation}
\textbf{F}=((000);(011);(022);(101);(112);(120);(202);(210);(221)).
\end{equation}
%  \begin{equation}
%\textbf{F}= \left(\begin{matrix}
%     0& 0& 0\cr
 %      0 & 1& 1\cr
 %   0 & 2& 2\cr
%   1 & 0& 1\cr
 %   1& 1& 2\cr
 %      1 & 2& 0\cr
%    2 & 0& 2\cr
%   2 & 1& 0\cr
%   2 & 2& 1\cr
%\end{matrix}\right)
%\end{equation}

In this paper, we only consider the OAs with all rows being different, and thus there will be no need to specifically distinguish an OA and the set of its row vectors.
\subsection{PDA Constructions via OAs}
It has been shown that row indices of PDAs can be represented by OAs \cite{29}. We briefly review the constructions of PDAs based on OAs \cite{16,13,19,29}.

 Given $q, m\in\mathbb{N}^{+}$ with $q\geq2$, let $\mathcal{F}=$OA$(m, q, m)$ and $\mathcal{K}=\{\textbf{k}=(\xi_0,c_0)\mid \xi_0\in[0,m], c_0\in [0,q-1]\}$. A $q^{m}\times (m+1)q$ array $\textbf{P}=(p_{\textbf{f},\textbf{k}})$, where $\textbf{f}=(f_0,f_1,\cdots,f_{m-1})\in\mathcal{F}$ and $\textbf{k}=(\xi_0,c_0)\in\mathcal{K}$, can be constructed with entries $p_{\textbf{f},\textbf{k}}$ defined as

  When $\xi_0\in[0,m-1]$,
\begin{equation}
        p_{\textbf{f},\textbf{k}}= \begin{cases}
         (f_0,\cdots, c_0,\cdots, f_{m-1}, f_{\xi_0}-c_0-1),& \text{if $f_{\xi_0}\neq c_0$ }\\
         \ast, &\text{otherwise}.
 \end{cases}
\end{equation}

  When $\xi_0=m$,
\begin{equation}
        p_{\textbf{f},\textbf{k}}= \begin{cases}
         (f_0,f_1, \cdots, f_{m-1}, c_0-\sum\limits_{i=0}^{m-1}f_i-1),& \text{if $\sum\limits_{i=0}^{m-1}f_i\neq c_0$ }\\
         \ast, &\text{otherwise}.
 \end{cases}
\end{equation}

Note that modulo $q$ computations are performed with the construction above. The construction leads to the following result.

\emph{\textbf{Lemma 3}} \cite{13}. Given $q, m\in\mathbb{N}^{+}$ with
$q\geq 2$, there always exists an $((m+1)q$, $q^{m}$, $q^{m-1}$, $q^{m+1}-q^{m})$ PDA.
%which gives a $q^{m}$-division $((m+1)q, M, N)$ coded caching scheme with a memory ratio of $\frac{M}{N}=\frac{1}{q}$ and a transmission rate of $R=q-1$.

  The PDA construction of \cite{16} can be viewed as a generalization of (5) by considering $t\geq 1$. Given $q, m$$, t\in\mathbb{N}^{+}$ with $t < m$ and $q\geq2$, let $\mathcal{F}=$OA$(m, q, m)$ and $\mathcal{K}=$$\{\textbf{k}=$$(\{\xi_0,\xi_1,\cdots,\xi_{t-1}\},$\\$(c_0,c_1,\cdots,c_{t-1}))\mid \{\xi_0$, $\xi_1,\cdots, \xi_{t-1}\}\in {[0,m-1]\choose t}, \xi_0<\xi_1<\cdots<\xi_{t-1}, c_i\in [0,q-1], i\in[0, t-1]\}$. A $q^{m}\times {m\choose t}q^{t}$ array $\textbf{P}=(p_{\textbf{f},\textbf{k}})$, where $\textbf{f}=(f_0,f_1,\cdots,f_{m-1})\in\mathcal{F}$ and $\textbf{k}=(\{\xi_0,\xi_1,\cdots,\xi_{t-1}\},(c_0,c_1,\cdots,c_{t-1}))\in\mathcal{K}$, can be constructed with entries $p_{\textbf{f},\textbf{k}}$ defined as
\begin{equation}
        p_{\textbf{f},\textbf{k}}= \begin{cases}
         (f_0,\cdots, c_h,\cdots, f_{m-1}, f_{\xi_0}-c_0-1, \cdots, f_{\xi_{t-1}}-c_{t-1}-1),&\text{if $f_{\xi_h}\neq c_h$ for $h\in[0, t-1]$}\\
         \ast, &\text{otherwise}.
 \end{cases}
\end{equation} Based on the above construction, the following result can be obtained.

\emph{\textbf{Lemma 4}} \cite{16}. Given $q, m, t\in\mathbb{N}^{+}$ with
$q\geq 2$ and $t<m$, there always exists an $({m\choose t}q^{t}$, $q^{m}$, $q^{m}-q^{m-t}(q-1)^{t}$, $q^{m}(q-1)^{t})$ PDA.
 %which gives a $q^{m}$-division $({m\choose t}q^{t}, M, N)$ coded caching scheme with a memory ratio of $\frac{M}{N}=1-(\frac{q-1}{q})^{t}$ and a transmission rate of $R=(q-1)^{t}$.

The PDA construction of \cite{16} was later generalized in \cite{19} through changing its entry rule. Given $q, z, m, t\in\mathbb{N}^{+}$ with $z < q$, $q\geq2$ and $t < m$, let $\mathcal{F}=\{(\textbf{f},\textbf{g})=((f_0, f_1, \cdots, f_{m-1}),(g_0, g_1,$\\$\cdots, g_{t-1}))\mid(\textbf{f},\textbf{g})\in$ OA$(m, q, m)$$\times[0, \lfloor\frac{q-1}{q-z}\rfloor-1]^{t}\}$ and $\mathcal{K}=\{\textbf{k}=(\{\xi_0,\xi_1,\cdots,\xi_{t-1}\},(c_0,c_1,$\\$\cdots,c_{t-1}))\mid \{\xi_0$, $\xi_1,\cdots, \xi_{t-1}\}\in {[0,m-1]\choose t}, \xi_0<\xi_1<\cdots<\xi_{t-1}, c_i\in [0,q-1], i\in[0, t-1]\}$. A $\lfloor\frac{q-1}{q-z}\rfloor^{t}q^{m}\times {m\choose t}q^{t}$ array $\textbf{P}=(p_{(\textbf{f},\textbf{g}),\textbf{k}})$ can be constructed with entries $p_{(\textbf{f},\textbf{g}),\textbf{k}}$ defined as
\begin{equation}
        p_{(\textbf{f},\textbf{g}),\textbf{k}}= \begin{cases}
         (f_0,\cdots, c_h-g_h(q-z),\cdots, f_{m-1}, f_{\xi_0}-c_0-1, & \text{if $f_{\xi_h}\notin\{c_h, c_h-1, \cdots, c_h$}\\\cdots, f_{\xi_{t-1}}-c_{t-1}-1),& \text{ $-(z-1)\}$ for $h\in[0, t-1]$}\\
         \ast, &\text{otherwise}.
 \end{cases}
\end{equation} This construction leads to the following result with more flexible parameters.

\emph{\textbf{Lemma 5}} \cite{19}. Given $q, z, m, t\in\mathbb{N}^{+}$ with
$q\geq 2, z<q$ and $t<m$, there always exists an $({m\choose t}q^{t}$, $\lfloor\frac{q-1}{q-z}\rfloor^{t}q^{m}$, $\lfloor\frac{q-1}{q-z}\rfloor^{t}[q^{m}-q^{m-t}(q-z)^{t}]$, $q^{m}(q-z)^{t})$ PDA.
 %which gives a $\lfloor\frac{q-1}{q-z}\rfloor^{t}q^{m}$-division $({m\choose t}q^{t}, M, N)$ coded caching scheme with a memory ratio of $\frac{M}{N}=1-(\frac{q-z}{q})^{t}$ and a transmission rate of $R=\frac{(q-z)^{t}} {\lfloor\frac{q-1}{q-z}\rfloor^{t}}$.

The above PDA constructions were further improved in yielding a smaller subpacketization level in \cite{29}.
 Given $q, m, t\in\mathbb{N}^{+}$ with $q\geq2$ and $t < m$, let $\mathcal{F}=$ OA$(m,q,m-1)$ and $\mathcal{K}=\{(\{\xi_0,\xi_1,\cdots,\xi_{t-1}\},(c_0,c_1,\cdots,c_{t-1}))\mid \{\xi_0$, $\xi_1,\cdots, \xi_{t-1}\}\in {[0,m-1]\choose t}, \xi_0<\xi_1<\cdots<\xi_{t-1}, c_i\in [0,q-1], i\in[0, t-1]\}$. An $|\mathcal{F}|\times|\mathcal{K}|$ array $\textbf{P}=(p_{\textbf{f},\textbf{k}})$, where $\textbf{f}=(f_0,f_1,\cdots,f_{m-1})\in\mathcal{F}$ and $\textbf{k}=(\{\xi_0,\xi_1,\cdots,\xi_{t-1}\},(c_0,c_1,\cdots,c_{t-1}))\in\mathcal{K}$, can be obtained with entries $p_{\textbf{f},\textbf{k}}$ defined as
\begin{equation}
        p_{\textbf{f},\textbf{k}}= \begin{cases}
         (f_0,\cdots, c_h,\cdots, f_{m-1}, o((f_0,\cdots, c_h,\cdots, f_{m-1}))),& \text{if $f_{\xi_h}\neq c_h$ for $h\in[0, t-1]$}\\
         \ast, &\text{otherwise}.
 \end{cases}
\end{equation}
Note that $o((f_0,\cdots, c_h,\cdots, f_{m-1}))$ is the occurrence order of the vector $(f_0,\cdots, c_h,\cdots, f_{m-1})$ in column $\textbf{k}$. The construction leads to the following improved result.

\emph{\textbf{Lemma 6}} \cite{29}. Given $q, m, t\in\mathbb{N}^{+}$ with
$q\geq 2$ and $t<m$, there always exists an $({m\choose t}q^{t}$, $q^{m-1}$, $q^{m-1}-q^{m-t-1}(q-1)^{t}$, $q^{m-1}(q-1)^{t})$ PDA.
 %which gives a $q^{m-1}$-division $({m\choose t}q^{t}, M, N)$ coded caching scheme with a memory ratio of $\frac{M}{N}=1-(\frac{q-1}{q})^{t}$ and a transmission rate of $R=(q-1)^{t}$.
\section{The New PDA Schemes}
This section proposes some new PDA constructions via POAs. The designs can improve upon
some known schemes in either the subpacketization level or the transmission rate.
%Firstly, we propose a novel construction of PDA by selecting POAs, and the corresponding scheme can achieve a large coding gain and a lower subpacketization level. Secondly, based on the proposed PDA construction, we present an effective transformation that can derive a coded caching scheme with a smaller subpacketization level over the previous one. Finally, due to the fact that some $``\ast"$s of the proposed PDAs without generating multicasting opportunities in the delivery phase, we can delete such $``\ast"$s and also obtain some new schemes with good performance.
Before presenting the new PDA constructions, we first present the design intuition.
\subsection{Design Intuition}
Given a $(K, F, Z, S)$ PDA $\textbf{P}'$, if we replace $Z_0$ integer entries in each column of $\textbf{P}'$ with $``\ast"$s,
 then the resulting array is a $(K, F, Z+Z_0, S_0)$ PDA, denoted by $\textbf{P}_0$. Note that $S_0\leq S$ always holds. This implies that the transmission rate of the scheme based on $\textbf{P}_0$ may be smaller than the scheme based on $\textbf{P}'$. Furthermore, we prefer to design a PDA $\textbf{P}=(\textbf{P}_0; \textbf{P}_1)$ with the same memory ratio as $\textbf{P}_0$ by adding a well designed $F_0\times K$ array $\textbf{P}_1$ to $\textbf{P}_0$ without increasing $S_0$. This is because the new transmission rate $\frac{S_0}{F+F_0}$ is smaller than $\frac{S_0}{F}$. Let us present an example to illustrate the main idea of our construction.
 %When $q=5, m=2, t=1$, based on the construction of \cite{29}, the following PDA $(10,5,1,20)$ \textbf{P} can be obtained. More details about the construction can be referred to \cite{29}.
 Given the following PDA $(10,5,1,20)$ $\textbf{P}'$, replacing integers from $[10, 19]$ of $\textbf{P}'$ with $``\ast"$s, we can obtain a new array $\textbf{P}_0=(\textbf{P}_{0,0} \;\textbf{P}_{0,1})$ as follows.
 \begin{figure}[H]
\centering
\includegraphics[scale=0.216]{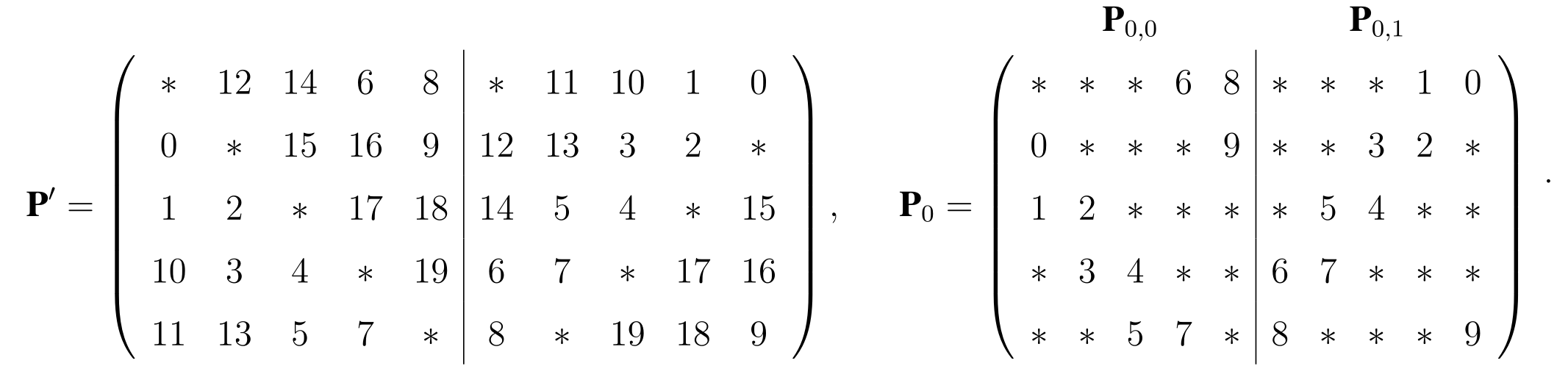}
%\caption{\label{2} The process of generating $\textbf{P}_2$ from $\textbf{P}_1$.}
 \end{figure}
Now we design an appropriate array $\textbf{P}_1$ by adjusting the integer entry of $\textbf{P}_0$ in Fig.2. Adding the well designed array $\textbf{P}_1$ to $\textbf{P}_0$ from top to bottom, a new $(10,5,6,10)$ PDA $\textbf{P}=(\textbf{P}_0; \textbf{P}_1)$ can be obtained as follows.
\begin{figure}[H]
\centering
\includegraphics[scale=0.295]{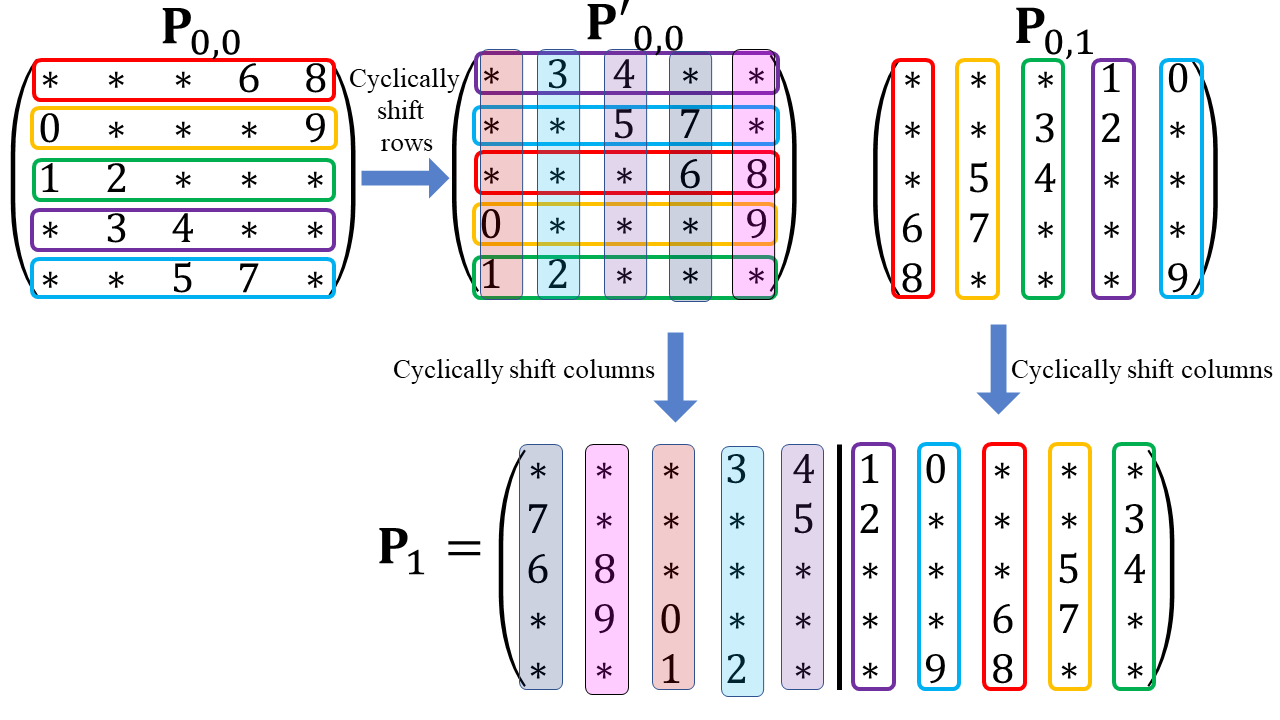}
\caption{\label{2}The process of generating $\textbf{P}_1$ from $\textbf{P}_0$: Firstly, we divide $\textbf{P}_0$ into two $5\times 5$ arrays, i.e., $\textbf{P}_0=(\textbf{P}_{0,0}\; \textbf{P}_{0,1})$. Then we cyclically shift each row vector of $\textbf{P}_{0,0}$ by two rows from top to bottom, which is denoted as $\textbf{P}'_{0,0}$. We further cyclically shift each column vector of $\textbf{P}'_{0,0}$ by two columns from left to right to get the left part of $\textbf{P}_1$. Finally, we cyclically shift each column vector of $\textbf{P}_{0,1}$ by two columns from left to right to get the right part of $\textbf{P}_1$.}
 \end{figure}

\begin{figure}[H]
\centering
\includegraphics[scale=0.22]{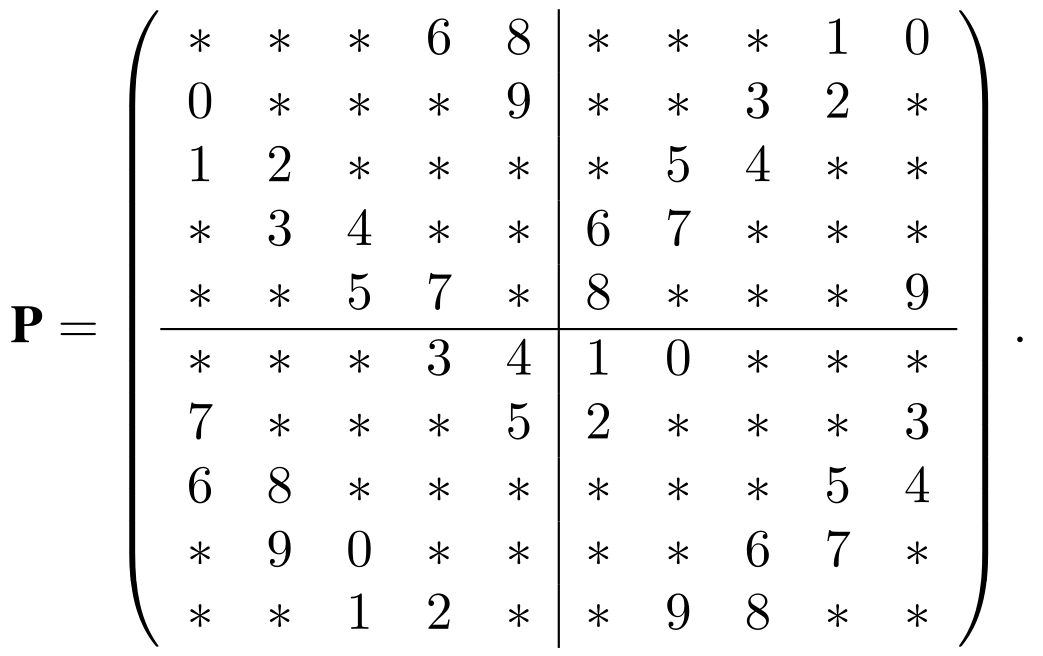}
%\caption{\label{2} The process of generating $\textbf{P}_2$ from $\textbf{P}_1$.}
 \end{figure}

 % \begin{equation}
%\textbf{P}=\left(
%  \begin{array}{c c c c c| c c c c c}
%   \ast  &  \ast   & \ast   &  6     & 8     &  \ast    & \ast  &  \ast  & 1     & 0 \\
%   0     &  \ast   & \ast   &  \ast  & 9     &  \ast    & \ast  &  3     & 2     & \ast\\
%   1     &  2      & \ast   &  \ast  & \ast  &  \ast    & 5     &  4     & \ast  & \ast\\
%   \ast  &  3      & 4      &  \ast  & \ast  &  6       & 7     &  \ast  & \ast  & \ast\\
%  \ast  &  \ast   & 5      &  7     & \ast  &  8       & \ast  &  \ast  & \ast  & 9\\\hline
%    \ast &  \ast   & \ast   &  3     & 4     &  1       & 0     &  \ast  & \ast  & \ast \\
%   7     &  \ast   & \ast   &  \ast  & 5     &  2       & \ast  &  \ast     & \ast  & 3\\
%   6     &  8      & \ast   &  \ast  & \ast  &  \ast    & \ast  &  \ast     & 5     & 4\\
%   \ast  &  9      & 0     &  \ast  & \ast  &  \ast    & \ast  &  6  & 7     & \ast\\
%   \ast  &  \ast   & 1      &  2     & \ast  &  \ast    & 9     &  8   & \ast  & \ast
%   \end{array}
%  \right).
%\end{equation}
  In general, a PDA $\textbf{P}=(\textbf{P}_0;\textbf{P}_1;\cdots;\textbf{P}_L)$ constructed by the above method can be viewed as
replacing the same number of integer entries with $``\ast"$s in each column of a given baseline array $\textbf{P}'$, and adding new arrays $\textbf{P}_1,\cdots,\textbf{P}_L$ with the same memory ratio as $\textbf{P}_0$ from top to bottom.
%as the one obtained by replacing some integer entries of $\textbf{P}'$ with $``\ast"$s.
In order to minimize the transmission rate of the scheme based on $\textbf{P}$, one needs to guarantee that if an integer entry $s$ occurring in some row and column of $\textbf{P}'$ is replaced by $``\ast"$, then all the entries of $\textbf{P}'$ containing $s$ are also replaced by $``\ast"$s as well. Furthermore, the newly added arrays $\textbf{P}_1,\cdots,\textbf{P}_L$ should be well designed such that their integer entries are the same with $\textbf{P}_0$. This implies that the main technical challenge for the above construction is how to design a baseline array $\textbf{P}'$ and newly added arrays $\textbf{P}_1,\cdots,\textbf{P}_L$ that can satisfy such constraints.

%Our main objective is to find an optimal baseline arrays and newly added arrays that can satisfy above constraints. Different from the construction of \cite{19}, we select
%Note that the PDA construction of \cite{19} is obtained by selection the PDA of \cite{16} as a baseline array. We intend to obtain a similar result as \cite{19} (i.e., the coding gain of each entry is ${m\choose t} \lfloor\frac{q-1}{q-z}\rfloor^{t}$) but a smaller subpacketization level
 \textbf{Note}: We should point out that our proposed method of constructing PDA is different with the construction of \cite{19}. In the construction of \cite{19}, all the row indices of newly added arrays are generated by the same OA$(m,q,m-1)$ and their integer entries are obtained by moving the entries of a designed array in a counter-clockwise.
Furthermore, if we directly apply the similar technique in \cite{19} to that in \cite{29}, it will violate above constraints. Therefore, in order to obtain a PDA with the coding gain of each entry as large as possible (i.e., the transmission rate as small as possible), some more empirical insights and technical methods should be utilized to the construction so that the above constraints can be satisfied.

\subsection{The New PDA Construction}
Based on the above observation, this subsection proposes a novel framework of constructing PDA.
%Based on the above observation, in the following subsection we propose a new general construction method of PDAs.
 It can be considered as a generalization of the work of \cite{29}. However, it requires a more delicate selection of the row indices of arrays.
%but it highly depends on smartly selecting the row indices of arrays.
Unlike the construction of \cite{19}, i.e., the row indices of PDA are obtained by reusing the row indices of the PDA in \cite{16} $\lfloor\frac{q-1}{q-z}\rfloor^{t}$ times, our construction is realized by carefully selecting the POAs. Now let us present some useful notations to our construction.

Given $q, z, m, t\in\mathbb{N}^{+}$ with $z < q$ and $t < m$, the packet indices and users can be represented as follows.

$\bullet$ Let $\mathcal{E}=\{(g_0, g_1, \cdots, g_{t-1})\mid g_i\in [0, \lfloor\frac{q-1}{q-z}\rfloor-1],i\in[0, t-1]\}$ and let
\begin{equation}
\begin{aligned}
&\textbf{F}_{\textbf{g}_j}=(\textbf{f}_{0}^{(j)}; \textbf{f}_{1}^{(j)}; \cdots; \textbf{f}_{q^{m-1}-1}^{(j)})\\
&=(( f_{0,0}^{(j)}, f_{0,1}^{(j)},\cdots, f_{0,m-1}^{(j)});(f_{1,0}^{(j)}, f_{1,1}^{(j)},\cdots, f_{1,m-1}^{(j)});\cdots;
(f_{q^{m-1}-1,0}^{(j)}, f_{q^{m-1}-1,1}^{(j)},\cdots, f_{q^{m-1}-1,m-1}^{(j)}))
\end{aligned}
\end{equation}
%\begin{equation}
%\textbf{F}_{\textbf{g}_j}=\left(\begin{matrix}
%     \textbf{f}_{0}^{(j)}\\
%     \textbf{f}_{1}^{(j)}\\
%  \vdots\\
 %  \textbf{f}_{q^{m-1}-1}^{(j)}
 %  \end{matrix}\right)
%=\left(\begin{matrix}
%     f_{0,0}^{(j)}        & f_{0,1}^{(j)}        &\cdots     & f_{0,m-1}^{(j)} \cr
%    f_{1,0}^{(j)}        & f_{1,1}^{(j)}        &\cdots     & f_{1,m-1}^{(j)} \cr
 % \vdots    &\vdots   &\cdots     &\vdots\cr
 %   f_{q^{m-1}-1,0}^{(j)}        & f_{q^{m-1}-1,1}^{(j)}        &\cdots     & f_{q^{m-1}-1,m-1}^{(j)} \cr
%\end{matrix}\right)
%\end{equation}\\
 denote a POA$(m,q,m-1)$ such that $\sum_{r=0}^{m-1}f_{s,r}^{(j)}=x(q-z)$ for $s\in[0,q^{m-1}-1]$ with $x=\sum_{i=0}^{t-1}g^{(j)}_i$, where $\textbf{g}_j=(g^{(j)}_0, g^{(j)}_1, \cdots, g^{(j)}_{t-1})\in\mathcal{E}$ and $\mathcal{E}=\{\textbf{g}_0,\textbf{g}_1,\cdots,\textbf{g}_{\lfloor\frac{q-1}{q-z}\rfloor^{t}-1}\}$.
 Each file is split into $F$ packets of equal size, which are labelled by $W_n=\{W_{n,(\textbf{f}^{(j)}_s,\textbf{g}_j)}\mid  s\in[0,q^{m-1}-1],j\in[0,\lfloor\frac{q-1}{q-z}\rfloor^{t}-1]\}$, where $\textbf{f}^{(j)}_s\in \textbf{F}_{\textbf{g}_j}$ and $n\in[0, N-1]$.

$\bullet$ Let $\mathscr{I}=\{\{\xi_0, \xi_1,\cdots, \xi_{t-1}\}\mid \{\xi_0$, $\xi_1,\cdots, \xi_{t-1}\}\in {[0,m-1]\choose t}$, $0\leq \xi_0<\xi_1< \cdots<\xi_{t-1}<m\}$ and $\mathcal{C}=\{(c_0, c_1,\cdots, c_{t-1})\mid c_i\in [0,q-1], i\in[0, t-1]\}$. Each user can be represented by a unique set pair $(\mathcal{I}, \textbf{c})$, where $\mathcal{I}\in \mathscr{I}$ and $\textbf{c}\in\mathcal{C}$.

Armed with the above notations, the new PDA construction can be proposed as follows.

\emph{\textbf{Construction 1}}. Given $q, z, m, t\in\mathbb{N}^{+}$ with $z < q$ and $t < m$, let
$\mathcal{K}=\{(\mathcal{I}, \textbf{c})=(\{\xi_0, \xi_1, \cdots,$\\$ \xi_{t-1}\}, (c_0,\cdots, c_{t-1}))\mid \mathcal{I}\in \mathscr{I}, \textbf{c}\in \mathcal{C}\}$
%$$\mathcal{F}=\{(\textbf{f}, \textbf{g})=((a_0, \cdots, a_{m-1}), (\varepsilon_0, \cdots, \varepsilon_{t-1}))\mid \textbf{f}\in \mathcal{M}, \textbf{g}\in \mathcal{E}\}$$
and
$\mathcal{F}=\mathop\bigcup_{j\in[0,\lfloor\frac{q-1}{q-z}\rfloor^{t}-1]}\mathcal{F}^{(j)}_{\textbf{g}_j},$
where $ \mathcal{F}^{(j)}_{\textbf{g}_j}=\{(\textbf{f}^{(j)}_{s},\textbf{g}_j)=((f^{(j)}_{s,0}, \cdots, f^{(j)}_{s,m-1}), (g^{(j)}_0, \cdots, g^{(j)}_{t-1}))\mid  s\in[0,q^{m-1}-1]\}$ and $\textbf{f}^{(j)}_{s}\in\textbf{F}_{\textbf{g}_j}$.
An $F\times K$ array
$\textbf{P}=(
     \textbf{P}_{0};$\\$
     \cdots;
     \textbf{P}_{j};
     \cdots;
     \textbf{P}_{\lfloor\frac{q-1}{q-z}\rfloor^{t}-1}
   )$
can be constructed with the entries of
$\textbf{P}_j=(p_{(\textbf{f}_s^{(j)}, \textbf{g}_j),(\mathcal{I}, \textbf{c})})_{
(\textbf{f}^{(j)}_{s},\textbf{g}_j)\in\mathcal{F}^{(j)}_{\textbf{g}_j}, (\mathcal{I},\textbf{c})\in\mathcal{K}}$ defined as
\begin{equation}
        p_{{(\textbf{f}^{(j)}_{s}, \textbf{g}_j),(\mathcal{I}, \textbf{c})}}= \begin{cases}
         (\textbf{v}, o(\textbf{v})), & \text{if $f_{s,\xi_h}^{(j)}\notin\{c_h, c_h-1, \cdots, c_h$}\\& \text{ $-(z-1)\}$ for $h\in[0, t-1]$}\\
         \ast, &\text{otherwise},
 \end{cases}
\end{equation} where $\textbf{v}=(v_0, v_1, \cdots, v_{m-1})\in[0, q-1]^{m}$ such that
\begin{equation}
        v_i= \begin{cases}
         c_h-g^{(j)}_h(q-z), & \text{if $i=\xi_h, h\in[0, t-1]$}\\
         f^{(j)}_{s,i}, &\text{otherwise}.
 \end{cases}
\end{equation} Note that $o(\textbf{v})$ is the occurrence order of vector $\textbf{v}$ in column $(\mathcal{I}, \textbf{c})$ and the computations are performed in modulo $q$.

The following example illustrates the above construction.
%\begin{equation}
%\textbf{F}_{(0)}
%   =\left(\begin{matrix}
%     0 & 0\cr
%    1 & 4\cr
%    2 & 3\cr
%     3 & 2\cr
 %      4 & 1\cr
%\end{matrix}\right),
%\textbf{F}_{(1)}
%=\left(\begin{matrix}
%     0 & 2\cr
 %   1 & 1\cr
 %   2 & 0\cr
  %   3 & 4\cr
  %     4 & 3\cr
%\end{matrix}\right).
%\end{equation}

\emph{\textbf{Example 1}}. Given $m=2$, $q=5$ and $t=1$, when $z=3$, we have $\lfloor\frac{q-1}{q-z}\rfloor=2$ and $\mathcal{E}=\{(0),$\\$(1)\}$. Let $\textbf{F}_{(0)}$ and $\textbf{F}_{(1)}$ denote two POA$(2,5,1)$s that are defined in (13). Note that the sum of
\begin{equation}
\textbf{F}_{(0)}=((00);(14);(23);(32);(41));\;\;\;\; \textbf{F}_{(1)}=((02);(11);(20);(34);(43)).
\end{equation}
  each row of $\textbf{F}_{(0)}$ is $0\times 2=0$ and the sum of each row of $\textbf{F}_{(1)}$ is $1\times2=2$. Hence,
%\begin{equation*}
%\textbf{F}_{(0)}
%  =\left(\begin{matrix}
  %   0 & 0\cr
 %   1 & 4\cr
 %   2 & 3\cr
 %    3 & 2\cr
 %      4 & 1\cr
%\end{matrix}\right),
%\textbf{F}_{(1)}
%=\left(\begin{matrix}
%     0 & 2\cr
%    1 & 1\cr
 %   2 & 0\cr
 %    3 & 4\cr
 %      4 & 3\cr
%\end{matrix}\right).
%\end{equation*}
 we have
$$\mathcal{F}=(\textbf{F}_{(0)}\times \{(0)\})\bigcup(\textbf{F}_{(1)}\times \{(1)\});\;\;\;\;
 \mathcal{K}=\{(\{\xi_0\},(c_0))\mid \xi_0\in[0,1], c_0\in [0,4]\}.$$
For $(\{\xi_0\},(c_0))=(\{0\},(0))\in\mathcal{K}$, we have $\{c_0, c_0-1, \cdots, c_0-(z-1)\}=\{0,3,4\}$. Based on\\ (11) and (12), for any $((f^{(j)}_{s,0},f^{(j)}_{s,1}),(g^{(j)}_0))\in\mathcal{F}$, we have $p_{((f^{(j)}_{s,0},f^{(j)}_{s,1}),(g^{(j)}_0)),(\{0\},(0))}=\ast$, if $f^{(j)}_{s,0}\in\{0,$\\$3,4\}$, and $p_{((f^{(j)}_{s,0},f^{(j)}_{s,1}),(g^{(j)}_0)),(\{0\},(0))}=(-2g^{(j)}_0,f^{(j)}_{s,1})$, if $f^{(j)}_{s,0}\notin\{0,3,4\}$. Further, based on (11), if $f^{(j)}_{s,0}\notin\{0,3,4\}$, $p_{((f^{(j)}_{s,0},f^{(j)}_{s,1}),(g^{(j)}_0)),(\{0\},(0))}=(-2g^{(j)}_0,f^{(j)}_{s,1}, o((-2g^{(j)}_0,f^{(j)}_{s,1}))$. E.g., since (04) first occurs in column $(\{0\},(0))$, and with the assumption that the occurrence order starts from 0, it can be seen that $p_{{((14),(0)),(\{0\},(0))}}=(040)$. Similarly, we can obtain $p_{((f^{(j)}_{s,0},f^{(j)}_{s,1}),(g^{(j)}_0)),(\{\xi_0\},(c_0))}$ for any $((f^{(j)}_{s,0},f^{(j)}_{s,1}),(g^{(j)}_0))\in\mathcal{F}, (\{\xi_0\},(c_0))\in\mathcal{K}$. As a result, the following PDA $\textbf{P}=(\textbf{P}_0;\textbf{P}_1)$ can be obtained.
\begin{center}
\includegraphics[scale=0.22]{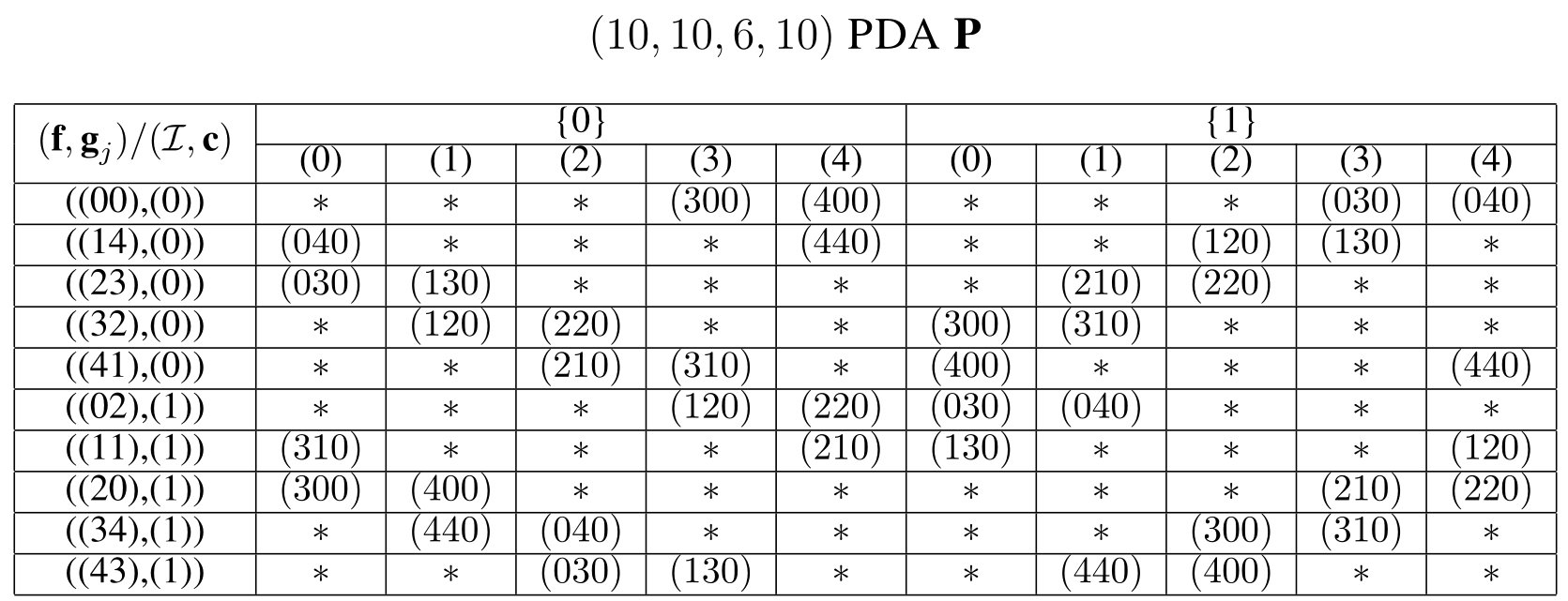}
\end{center}
Based on {\it Construction 1}, the following result presents a new class of PDAs that can yield a coded caching scheme with a smaller subpacketization level than the one in \cite{19}. For the sake of presentation consistency, its proof is given in Appendix A.

\emph{\textbf{Theorem 1}}. Given $q, z, m, t\in\mathbb{N}^{+}$ with
$q\geq 2, z<q$ and $t<m$, there always exists an $({m\choose t}q^{t}$, $\lfloor\frac{q-1}{q-z}\rfloor^{t}q^{m-1}$, $\lfloor\frac{q-1}{q-z}\rfloor^{t}[q^{m-1}-q^{m-t-1}(q-z)^{t}]$, $q^{m-1}(q-z)^{t})$ PDA which yields a $\lfloor\frac{q-1}{q-z}\rfloor^{t}q^{m-1}$-division $({m\choose t}q^{t}, M, N)$ coded
caching scheme with a memory ratio of $\frac{M}{N}=1-(\frac{q-z}{q})^{t}$ and a transmission rate of $R=\frac{(q-z)^{t}} {\lfloor\frac{q-1}{q-z}\rfloor^{t}}$.

%\emph{\textbf{Remark} 1}.
 %Existing constructions and theoretical analysis imply that to construct a coded caching scheme with smaller subpacketization level, one may sacrifice some rates. However, our proposed scheme in Theorem 1 can yield a smaller subpacketization, while maintains the same transmission rate, i.e.,
%Given $q, z, m, t\in\mathbb{N}^{+}$ with $q\geq 2, z<q$ and $t<m$, our scheme in Theorem 1 has the same number of users, memory ratio and transmission rate as the first scheme in \cite{19}, while our subpacketization is only $\frac{1}{q}$ of the first scheme in \cite{19}.
%When $z=1$, our scheme in  Theorem 1 is the same as the scheme in \cite{29}. However, our scheme generalizes the scheme in \cite{29} with a wider range of applications than the original one due to a more flexible memory size.

\subsection{The Generic Transformation}
Inspired by {\it Construction 1}, we now introduce an effective transformation of {\it Construction 1} that can derive a coded caching scheme with a smaller subpacketization level than the scheme in {\it Theorem 1}. The transformation can be performed through the following two steps. %For consistency, we borrow the notations in Construction 1.

\textbf{Step 1:} Choose the new $\hat{\textbf{P}}_0, \hat{\textbf{P}}_1, \cdots, \hat{\textbf{P}}_{\lfloor\frac{q-1}{q-z}\rfloor^{t}-1}$ as the baseline arrays

%For convenience, we partition the set $\mathcal{E}=\{\textbf{g}_0,\textbf{g}_1,\cdots,\textbf{g}_{\lfloor\frac{q-1}{q-z}\rfloor^{t}-1}\}$ into several disjoint subsets, i.e., $\mathcal{E}=\mathcal{E}_0\cup\mathcal{E}_1\cup\cdots\cup\mathcal{E}_{t(\lfloor\frac{q-1}{q-z}\rfloor-1)}$, where $\mathcal{E}_0,\mathcal{E}_1,\cdots,\mathcal{E}_{t(\lfloor\frac{q-1}{q-z}\rfloor-1)}$ are defined as
%\begin{align*}
%&\mathcal{E}_0=\{(0,0,\cdots,0)\};\\
%&\mathcal{E}_1=\{\textbf{g}+\textbf{x}_i\mid \textbf{g}\in \mathcal{E}_0, i\in[0,t-1]\}\backslash \mathcal{E}_0;\\
%&\mathcal{E}_2=\{\textbf{g}+\textbf{x}_i\mid \textbf{g}\in \mathcal{E}_1, i\in[0,t-1]\}\backslash (\mathcal{E}_0\cup\mathcal{E}_1);\\
%&\mathcal{E}_3=\{\textbf{g}+\textbf{x}_i\mid \textbf{g}\in \mathcal{E}_2, i\in[0,t-1]\}\backslash (\mathcal{E}_0\cup\mathcal{E}_1\cup\mathcal{E}_2);\\
%&\vdots\\
%&\mathcal{E}_{t(\lfloor\frac{q-1}{q-z}\rfloor-1)}=\{\textbf{g}+\textbf{x}_i\mid \textbf{g}\in \mathcal{E}_{t(\lfloor\frac{q-1}{q-z}\rfloor-1)-1}, i\in[0,t-1]\}\backslash (\mathcal{E}_0\cup\cdots\cup\mathcal{E}_{t(\lfloor\frac{q-1}{q-z}\rfloor-1)-1});
%\end{align*}
%and $\textbf{x}_i=(0,\cdots,1,\cdots,0)$ is a vector with the $i$th entry being 1 and others 0s.
%The computations are performed in modulo $\lfloor\frac{q-1}{q-z}\rfloor$.
Let $\textbf{P}=(
     \textbf{P}_{0};
     \cdots;
     \textbf{P}_{j};
     \cdots;
     \textbf{P}_{\lfloor\frac{q-1}{q-z}\rfloor^{t}-1}
   )$
%\begin{equation*}
%\textbf{P}=\left(\begin{matrix}
%     \textbf{P}_{0}\\
%     \vdots\\
  %   \textbf{P}_{j}\\
  %   \vdots\\
  %   \textbf{P}_{\lfloor\frac{q-1}{q-z}\rfloor^{t}-1}
 %  \end{matrix}\right)
 %  \end{equation*}
 denote an array generated from {\it Construction 1} with the row indices of $\textbf{P}_j$ indexed by $\mathcal{F}_{\textbf{g}_j}^{(j)}$, where the first $m-1$ coordinates of the row indices of $\textbf{P}_j$ (the first $m-1$ coordinates of the rows of $\textbf{F}_{\textbf{g}_j}$) are arranged in the lexicographic order from top to bottom.
For any $j\in[1, \lfloor\frac{q-1}{q-z}\rfloor^{t}-1]$, array $\textbf{P}_j'$ can be obtained from $\textbf{P}_j$ as follows.

$\bullet$ Let $\textbf{P}_0'=\textbf{P}_0$.
%Let $\textbf{g}_0=(0,0,\cdots,0)\in\mathcal{E}$.
%It can be seen that $\mathcal{E}_1=\{\textbf{g}_0+\textbf{x}_i\mid i\in[0,t-1]\}$.
Given any $\textbf{g}_j=(g^{(j)}_0, g^{(j)}_1, \cdots, g^{(j)}_{t-1})\in\mathcal{E}$ and $\textbf{g}_0=(0,0,\cdots,0)$, if an entry $\textbf{e}=(\textbf{v},o(\textbf{v}))$ appears in row $((f^{(0)}_{s0},f^{(0)}_{s1},\cdots,f^{(0)}_{s\xi_i},\cdots,f^{(0)}_{sm-1}),\textbf{g}_0)$ and column $(\mathcal{I},\textbf{c})=(\{\xi_0,\xi_1,\cdots,\xi_{t-1}\},\textbf{c})$ of $\textbf{P}_0$, then the entry of $\textbf{P}_j$ that contains \textbf{v} in row $((f^{(0)}_{s0},\cdots,f^{(0)}_{s\xi_0}+g_0^{(j)}(q-z),\cdots,f^{(0)}_{s\xi_1}+g_1^{(j)}(q-z), \cdots,f^{(0)}_{s\xi_{t-1}}+g_{t-1}^{(j)}(q-z),\cdots,f^{(0)}_{sm-1}),\textbf{g}_j)$ and column $(\mathcal{I},\textbf{c}')$ is modified as $\textbf{e}$. Using this manner to all entries of $\textbf{P}_0$, we can obtain a new array $\textbf{P}_{j}'$ by modifying the entries of $\textbf{P}_{j}$.

$\bullet$ Let $\hat{\textbf{P}}_0=\textbf{P}_0'$.  For any $j\in[1,\lfloor\frac{q-1}{q-z}\rfloor^{t}-1]$, the baseline array $\hat{\textbf{P}}_j$ is obtained by selecting the columns of $\textbf{P}_j'$ indexed by $(\mathcal{I}, \textbf{c})$, where $\mathcal{I}\in {[0,m-2]\choose t}$ and $\textbf{c}\in\{(c_0, c_1,\cdots, c_{t-1})\mid c_i\in [0,q-1], i\in[0, t-1]\}$.

\textbf{Step 2:} Transform the baseline arrays into the new PDA $\hat{\textbf{P}}$

This transform is performed by arranging the baseline arrays from left to right as
$\hat{\textbf{P}}=(\hat{\textbf{P}}_{0}\;
     %\hat{\textbf{P}}_{1}\;
     \cdots\;
     \hat{\textbf{P}}_{j}\;
     \cdots\;
     \hat{\textbf{P}}_{\lfloor\frac{q-1}{q-z}\rfloor^{t}-1}).$
A new array $\hat{\textbf{P}}$ can be obtained with the number of columns being greater than that of \textbf{P}, while maintaining the same number of rows as $\textbf{P}_0$. By applying the transform to {\it Construction 1}, we have the following result, whose proof can be found in Appendix B.

\emph{\textbf{Theorem 2}}. Given $q, z, m, t\in\mathbb{N}^{+}$ with
$q\geq 2, z<q$ and $t<m$, there always exists an $([{m-1\choose t}\lfloor\frac{q-1}{q-z}\rfloor^{t}+{m\choose t}-{m-1\choose t}]q^{t}$, $q^{m-1}$, $q^{m-1}-q^{m-t-1}(q-z)^{t}$, $q^{m-1}(q-z)^{t})$ PDA which yields a $q^{m-1}$-division $([{m-1\choose t}\lfloor\frac{q-1}{q-z}\rfloor^{t}+{m\choose t}-{m-1\choose t}]q^{t}, M, N)$ coded caching scheme with a memory ratio of $\frac{M}{N}=1-(\frac{q-z}{q})^{t}$ and a transmission rate of $R=(q-z)^{t}$.

%\emph{\textbf{Remark} 2}. It can be seen that when the memory ratio is less than $\frac{1}{2}$, the scheme in Theorem 2 is the same as Theorem 1. However, when the memory ratio is larger than $\frac{1}{2}$, the scheme in Theorem 2 has a smaller subpacketization level and larger number of users at the cost of some transmission rates compared with Theorem 1.

The following example illustrates the above transform and its effect.

\emph{\textbf{Example 2}}. Let us consider the array
$\textbf{P}=(
     \textbf{P}_{0};
     \textbf{P}_{1})$
  of {\it Example 1}. Based on the above transform, we have $\hat{\textbf{P}}_{0}=\textbf{P}_{0}$. Furthermore, based on $\textbf{P}_{0}$, array $\textbf{P}_{1}'$ can be obtained from $\textbf{P}_{1}$ by modifying its entries. E.g., given $\textbf{g}_1=(1)\in\mathcal{E}_1$, if the entry (040) occurs in row $((14),(0))$ and column $\{\{0\},(0)\}$ of $\textbf{P}_0$, then the entry of $\textbf{P}_1$ contains vector (04) in row $((34),(1))$ and column $\{\{0\},(2)\}$ is modified as
 (040) since $q-z=2$. Similarly, the other entries of $\textbf{P}_{1}'$ can be obtained in the same manner.
 Then $\hat{\textbf{P}}_{1}$ can be obtained by selecting the columns of $\textbf{P}_{1}'$ indexed by $(\{0\}, (0)), (\{0\}, (1)), (\{0\}, (2)), (\{0\}, (3))$ and $(\{0\}, (4))$. Consequently, we have the following $(15, 5, 3, 10)$ PDA $\hat{\textbf{P}}=(\hat{\textbf{P}}_{0}\;\hat{\textbf{P}}_{1})$ which yields a $5$-division $(15, M, N)$ coded caching scheme with a memory ratio of $\frac{M}{N}=\frac{3}{5}$ and a transmission rate of $R=2$.
%\begin{center}
%\includegraphics[scale=0.20]{28.png}
%\end{center}
%\begin{center}
%\includegraphics[scale=0.20]{29.png}
%\end{center}
\begin{center}
\includegraphics[scale=0.21]{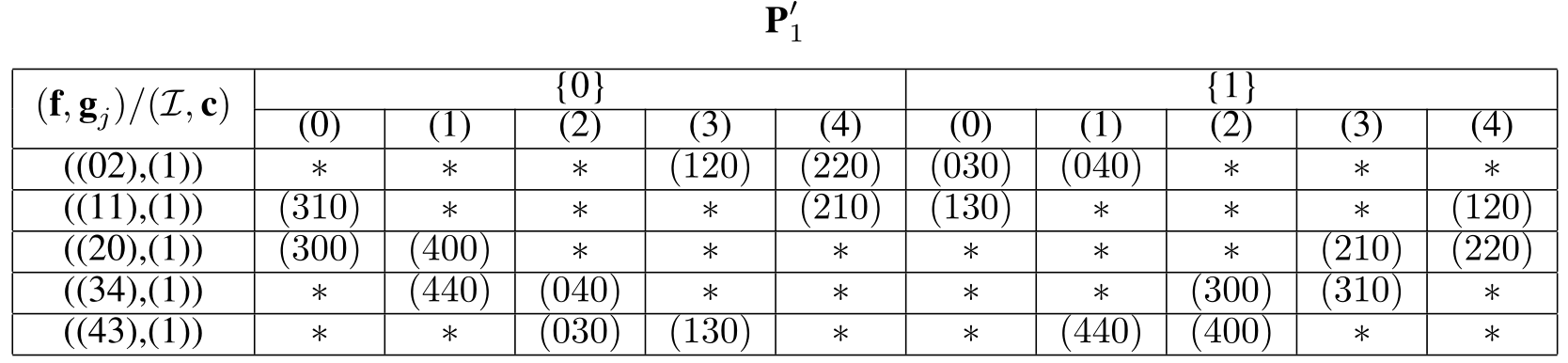}
\end{center}
\begin{center}
\includegraphics[scale=0.21]{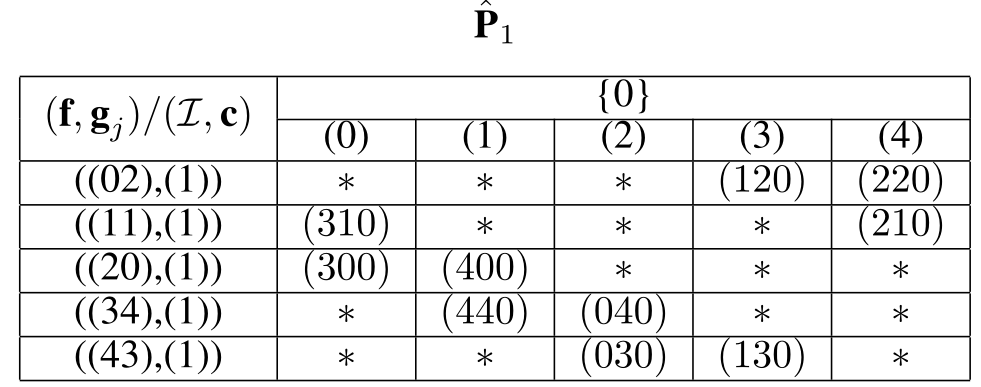}
\end{center}
\begin{center}
\includegraphics[scale=0.251]{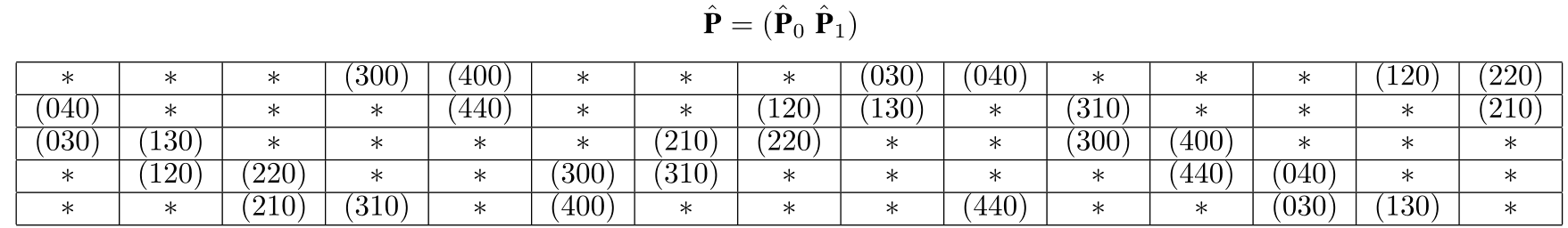}
\end{center}
%\renewcommand{\arraystretch}{1.3}
%\begin{table*}[!htp]
%\caption{$\hat{\textbf{P}}=(\hat{\textbf{P}}_0\;\hat{\textbf{P}}_1)$}
%\centering
% \fontsize{10}{8}\selectfont
%  \begin{tabular}{|c|c|c|c|c|c|c|c|c|c|c|c|c|c|c|}
%  \hline
%  $\ast$&$\ast$&$\ast$&$(300)$&$(400)$&$\ast$&$\ast$&$\ast$&$(030)$&$(040)$&$\ast$&$\ast$&$\ast$&$(120)$&$(220)$\cr\hline
% $(040)$&$\ast$&$\ast$&$\ast$&$(440)$&$\ast$&$\ast$&$(120)$&$(130)$&$\ast$&$(310)$&$\ast$&$\ast$&$\ast$&$(210)$\cr\hline
%$(030)$&$(130)$&$\ast$&$\ast$&$\ast$&$\ast$&$(210)$&$(220)$&$\ast$&$\ast$&$(300)$&$(400)$&$\ast$&$\ast$&$\ast$\cr\hline
%    $\ast$&$(120)$&$(220)$&$\ast$&$\ast$&$(300)$&$(310)$&$\ast$&$\ast$&$\ast$&$\ast$&$(440)$&$(040)$&$\ast$&$\ast$\cr\hline
%  $\ast$&$\ast$&$(210)$&$(310)$&$\ast$&$(400)$&$\ast$&$\ast$&$\ast$&$(440)$&$\ast$&$\ast$&$(030)$&$(130)$&$\ast$\cr\hline
% \end{tabular}
%\end{table*}
\subsection{New PDA Schemes with Coded Placement}
 Based on the above PDAs, two new coded caching schemes with the coded placement will be proposed. In a PDA, a $``\ast"$ is useless, if it is not contained in any subarray shown in C3-(b) of {\it Definition 1}. This indicates the useless $``\ast"$s cannot generate multicasting opportunities in the delivery phase, i.e., these $``\ast"$s have no advantage in reducing the transmission rate of a coded caching scheme realized by that PDA and result in a high subpacketization level. Therefore, if each column of a $(K, F, Z, S)$ PDA has $Z'$ useless $``\ast"$s, a new coded caching scheme with a smaller memory ratio and subpacketization level can be obtained by deleting these useless $``\ast"$s, and using an $[F, F-Z']_{q'}$ maximum distance separable code that is defined over $\mathbb{F}_{q'}$ in the placement phase, where $F$ and $F-Z'$ denote the length and the dimension of the code, respectively. More details can be referred to \cite{24}.

\emph{\textbf{Lemma 7}} \cite{24}. For any $(K, F, Z, S)$ PDA $\textbf{P}$, if there exist $Z'$
useless $``\ast"$s in each column, then we can obtain an $(F-Z')$-division $(K, M, N)$ coded caching scheme with a memory ratio of $\frac{M}{N}=\frac{Z-Z'}{F-Z'}$
and a transmission rate of  $R =\frac{S}{F-Z'}$, in which the coding gain at each time slot is the same as the original scheme realized by
$\textbf{P}$ and Algorithm 1.
%\begin{proof}
%Given a $(K,F,Z,S)$ PDA \textbf{P} with $Z'$ useless stars in each column, we can obtain a new array $\mathbf{P}'=(p'_{j,k}), j\in[0,F-1], k\in[0,K-1]$ by deleting these useless stars of \textbf{P}. It can be seen that each column of $\mathbf{P}'$ has $Z'$ blanks, $F-Z$ integers and $Z-Z'$ stars. Then based on $\mathbf{P}'$, the placement strategy in Algorithm 1 can be modified in the following way. Each file is divided into $F-Z'$ packets with equal size and then encodes them using an $[F,F-Z']_{q'}$ maximum distance separable (MDS) code for some prime power $q'$. Let the resulting encoded packets denote as $W_{n,0}$, $\cdots$, $W_{n,F-1}$ for each file $W_{n}, n\in[0,n-1]$. Using the caching strategy in Lines 3-4 of Algorithm 1, each user $k$ caches $Z_{k}=$$\{W_{n,j}$$|p'_{j,k}$=$\ast$, $j$$\in[0,F-1]$, $n\in[0,N-1]\}$. It can be seen that the memory ratio of each user is $\frac{M}{N}$$=$$\frac{Z-Z'}{F-Z'}$. In the delivery phase, we also use the delivery strategy in Algorithm 1 as follows. For any request vector $\textbf{d}$, using Lines 6-7 of Algorithm 1, each user can get exactly $F-Z'$ required coded packets. The property of $[F,F-Z']_{q'}$ MDS code ensures that each user can recover its requested file after receiving $F-Z'$ coded symbols with its cache help. So the transmission rate is $R=\frac{S}{F-Z'}$.
%\end{proof}

\emph{\textbf{Remark} 1}. Note that the operation field $q'$ of {\it Lemma 7} must be $O(F)$. Therefore, the size of each packet of files must approximate to $\log_2F$ bits. This implies that the size of files in the server must be more than $(F-Z')\log_2F$ so that the transmission rate of $\frac{S}{F-Z'}$ can be maintained. Given a $(K,F,Z,S)$ PDA, $\frac{Z}{F}>\frac{Z-Z'}{F-Z'}$ and $F>F-Z'$ always hold for any $Z,F,Z'\in\mathbb{N}^{+}$. This also implies that we can obtain a coded caching scheme with a smaller subpacketization level and memory ratio by deleting the useless stars.

Note that there exist some useless $``\ast"$s in each column of the array \textbf{P} generated from {\it Construction 1}. Therefore, based on {\it Lemma 7}, the following result can be obtained.
%the scheme in {\it Theorem 1} can be further improved with a smaller memory ratio and subpacketization level by using {\it Lemma 7}.
%we can obtain a coded caching scheme with a smaller memory ratio and subpacketization level, compared with Theorem 1 by using Lemma 7.

\emph{\textbf{Theorem 3}}. Given any $q, m, t\in\mathbb{N}^{+}$ with
$q\geq 2$ and $t<m$, let $z_{r}^{\ast}$ denote the minimal integer in the set $\mathcal{G}_r=\{z\mid \lfloor\frac{q-1}{q-z}\rfloor=r, z\in[1,q-1]\}$, where $r\in[1,q-1]$.
There exists an $({m\choose t}q^{t}, M, N)$ coded
caching scheme with a memory ratio of $\frac{M}{N}=\frac{1-(\frac{q-z_{r}^{\ast}}{q})^{t}}{1-(\frac{q-z_{r}^{\ast}}{q})^{t}+(\frac{q-z}{q})^{t}}$, a transmission rate of $R=\frac{(q-z)^{t}}{\lfloor\frac{q-1}{q-z}\rfloor^{t}(1-(\frac{q-z_{r}^{\ast}}{q})^{t}+(\frac{q-z}{q})^{t})}$, and a subpacketization level of $F=\lfloor\frac{q-1}{q-z}\rfloor^{t} q^{m-1}[1+(\frac{q-z}{q})^{t}-(\frac{q-z_{r}^{\ast}}{q})^{t}]$.
%\emph{\textbf{Theorem 3}}. Given $q, z, m, t\in\mathbb{N}^{+}$ with $q\geq 2, 2z\leq q$ and $t<m$, there exists $({m\choose t}q^{t}, M, N)$ a coded
%caching scheme with a memory ratio of $\frac{M}{N}\leq\frac{1-(\frac{q-1}{q})^{t}}{1-(\frac{q-1}{q})^{t}+(\frac{q-z}{q})^{t}}$, a transmission rate of $R\geq\frac{(q-z)^{t}}{1-(\frac{q-1}{q})^{t}+(\frac{q-z}{q})^{t}}$ and subpacketization $F\leq q^{m-1}[1+(\frac{q-z}{q})^{t}-(\frac{q-1}{q})^{t}]$.

\begin{proof}
Let \textbf{P} denote the PDA generated from {\it Construction 1}. When $z>z_{r}^{\ast}$ and $z\in\mathcal{G}_r$, array \textbf{P} can be viewed as replacing some vector entries with $``\ast"$s uniformly in the column where they occur from the case of $z=z_{r}^{\ast}$. This implies that there exist $\lfloor\frac{q-1}{q-z}\rfloor^{t}q^{m-1}[(\frac{q-z_{r}^{\ast}}{q})^{t}-(\frac{q-z}{q})^{t}]$ useless $``\ast"$s in each column of $\textbf{P}$. Furthermore, based on {\it Theorem 1}, the number of distinct vector entries in $\textbf{P}$ is $q^{m-1}(q-z)^{t}$. Combined with {\it Lemma 7}, the result holds.
\end{proof}

The following example illustrates the realization process of the scheme in {\it Theorem 3}.

\emph{\textbf{Example 3}}. Given $m=2$, $q=5$ and $t=1$,  based on {\it Construction 1}, when $z=1$, we obtain a PDA $\overline{\textbf{P}}_0$ as follows.
\begin{center}
\includegraphics[scale=0.225]{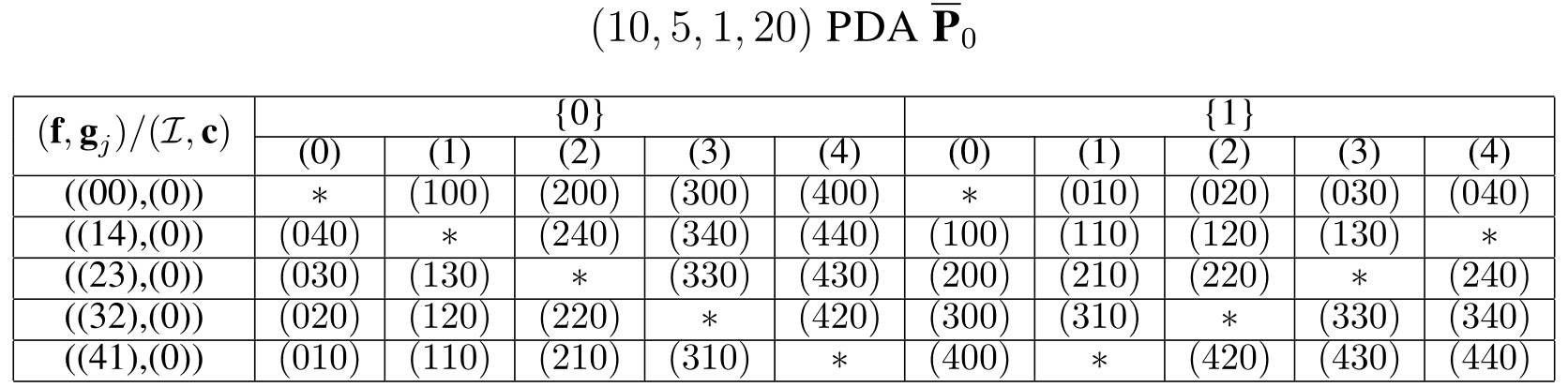}
\end{center}
%\renewcommand{\arraystretch}{1.3}
%\begin{table*}[!htp]

%\centering
% \fontsize{10}{8}\selectfont
%\caption{$(10,5,1,20)$ PDA $\overline{\textbf{P}}_4$}
%  \begin{tabular}{|c|c|c|c|c|c|c|c|c|c|c|}
%\hline
%   \multirow{2}{*}{$(\textbf{f}, \textbf{g}_j)/(\mathcal{I}, \textbf{c})$}&
 %  \multicolumn{5}{c|}{\{0\}}&\multicolumn{5}{c|}{ \{1\}}\cr\cline{2-11}
 % &(0)&(1)&(2)&(3)&(4)&(0)&(1)&(2)&(3)&(4)\cr
  %  \hline
  % ((00),(0))&$\ast$&$(100)$&$(200)$&$(300)$&$(400)$&$\ast$&$(010)$&$(020)$&$(030)$&$(040)$\cr\hline
  % ((14),(0))&$(040)$&$\ast$&$(240)$&$(340)$&$(440)$&$(100)$&$(110)$&$(120)$&$(130)$&$\ast$\cr\hline
 %((23),(0))&$(030)$&$(130)$&$\ast$&$(330)$&$(430)$&$(200)$&$(210)$&$(220)$&$\ast$&$(240)$\cr\hline
%    ((32),(0))&$(020)$&$(120)$&$(220)$&$\ast$&$(420)$&$(300)$&$(310)$&$\ast$&$(330)$&$(340)$\cr\hline
%  ((41),(0))&$(010)$&$(110)$&$(210)$&$(310)$&$\ast$&$(400)$&$\ast$&$(420)$&$(430)$&$(440)$\cr\hline

  %  ((02),(1))&$\ast$&$\ast$&$\ast$&$(120)$&$(220)$&$(030)$&$(040)$&$\ast$&$\ast$&$\ast$\cr\hline
 % ((11),(1))&$(310)$&$\ast$&$\ast$&$\ast$&$(210)$&$(130)$&$\ast$&$\ast$&$\ast$&$(120)$\cr\hline
  % ((20),(1))&$(300)$&$(400)$&$\ast$&$\ast$&$\ast$&$\ast$&$\ast$&$\ast$&$(210)$&$(220)$\cr\hline
   % ((34),(1))&$\ast$&$(440)$&$(040)$&$\ast$&$\ast$&$\ast$&$\ast$&$(300)$&$(310)$&$\ast$\cr\hline
   %((43),(1))&$\ast$&$\ast$&$(030)$&$(130)$&$\ast$&$\ast$&$(440)$&$(400)$&$\ast$&$\ast$\cr\hline
% \end{tabular}
%\end{table*}
If $r=1$, we have $\mathcal{G}_1=\{1,2\}$ and $z_{1}^{\ast}=1$. By taking $z=2$, the following PDA $\overline{\textbf{P}}_1$ can be obtained. It can be viewed as replacing some vector entries with $``\ast"$s uniformly in the columns where they occur from $\overline{\textbf{P}}_0$. The star with symbol $``\times"$ means that this star is useless and it can be deleted.
\begin{center}
\includegraphics[scale=0.225]{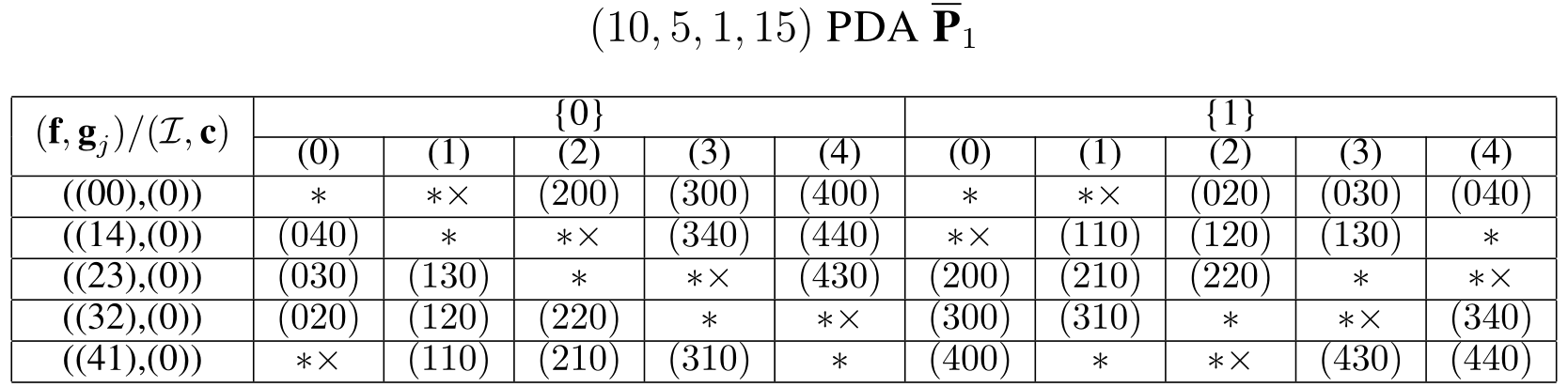}
\end{center}
%\renewcommand{\arraystretch}{1.3}
%\begin{table}[H]

%\centering
%\fontsize{10}{8}\selectfont
%\caption{$(10,5,1,15)$ PDA $\overline{\textbf{P}}_5$}
%   \begin{tabular}{|c|c|c|c|c|c|c|c|c|c|c|}
% \hline
 %  \multirow{2}{*}{$(\textbf{f}, \textbf{g}_j)/(\mathcal{I}, \textbf{c})$}&
 %  \multicolumn{5}{c|}{\{0\}}&\multicolumn{5}{c|}{ \{1\}}\cr\cline{2-11}
 % &(0)&(1)&(2)&(3)&(4)&(0)&(1)&(2)&(3)&(4)\cr
 %   \hline
 %  ((00),(0))&$\ast$&$\ast \times$&$(200)$&$(300)$&$(400)$&$\ast$&$\ast \times$&$(020)$&$(030)$&$(040)$\cr\hline
 %  ((14),(0))&$(040)$&$\ast$&$\ast \times$&$(340)$&$(440)$&$\ast \times$&$(110)$&$(120)$&$(130)$&$\ast$\cr\hline
% ((23),(0))&$(030)$&$(130)$&$\ast$&$\ast \times$&$(430)$&$(200)$&$(210)$&$(220)$&$\ast$&$\ast\times$\cr\hline
 %    ((32),(0))&$(020)$&$(120)$&$(220)$&$\ast$&$\ast \times$&$(300)$&$(310)$&$\ast$&$\ast \times$&$(340)$\cr\hline
 % ((41),(0))&$\ast \times$&$(110)$&$(210)$&$(310)$&$\ast$&$(400)$&$\ast$&$\ast \times$&$(430)$&$(440)$\cr\hline

  %  ((02),(1))&$\ast$&$\ast$&$\ast$&$(120)$&$(220)$&$(030)$&$(040)$&$\ast$&$\ast$&$\ast$\cr\hline
 % ((11),(1))&$(310)$&$\ast$&$\ast$&$\ast$&$(210)$&$(130)$&$\ast$&$\ast$&$\ast$&$(120)$\cr\hline
  % ((20),(1))&$(300)$&$(400)$&$\ast$&$\ast$&$\ast$&$\ast$&$\ast$&$\ast$&$(210)$&$(220)$\cr\hline
   % ((34),(1))&$\ast$&$(440)$&$(040)$&$\ast$&$\ast$&$\ast$&$\ast$&$(300)$&$(310)$&$\ast$\cr\hline
   %((43),(1))&$\ast$&$\ast$&$(030)$&$(130)$&$\ast$&$\ast$&$(440)$&$(400)$&$\ast$&$\ast$\cr\hline
% \end{tabular}
%\end{table}
Let $\overline{\textbf{P}}_2$ denote an array obtained by deleting the useless stars in $\overline{\textbf{P}}_1$. Based on {\it Lemma 7}, array $\overline{\textbf{P}}_2$ can realize a 4-division $(10, 2.5, 10)$ coded caching scheme as follows.

$\bullet$ \textbf{Placement Phase}: Each file $W_n$ is divided
into 4 packets, i.e., $W_n=\{W_{n,0}, W_{n,1}, W_{n,2}, W_{n,3}\}$ for $n\in[0, 9]$. Let $W_{n,4}=W_{n,0}+ W_{n,1}+ W_{n,2}+ W_{n,3}$.
Then the contents cached by the users are\begin{align*}
&\mathcal{Z}_0=\{W_{n,0}\mid n\in[0,9]\}; \mathcal{Z}_1=\{W_{n,1}\mid n\in[0,9]\};\mathcal{Z}_2=\{W_{n,2}\mid n\in[0,9]\};\\
&\mathcal{Z}_3=\{W_{n,3}\mid n\in[0,9]\};\mathcal{Z}_4=\{W_{n,4}\mid n\in[0,9]\};\mathcal{Z}_5=\{W_{n,0}\mid n\in[0,9]\};\\
&\mathcal{Z}_6=\{W_{n,4}\mid n\in[0,9]\};\mathcal{Z}_7=\{W_{n,3}\mid n\in[0,9]\};\mathcal{Z}_8=\{W_{n,2}\mid n\in[0,9]\};\\
&\mathcal{Z}_9=\{W_{n,1}\mid n\in[0,9]\}.
 \end{align*}

$\bullet$ \textbf{Delivery Phase}: Assume that the request vector is $\textbf{d}=(0, 1, 2, 3, 4, 5, 6, 7, 8, 9)$. The packets sent by the server at
fifteen time slots are listed as follows.
Time slot indexed by (040): $W_{0,1}\oplus W_{9,0}$; Time slot indexed by (030): $W_{0,2}\oplus W_{8,0}$; Time slot indexed by (020): $W_{0,3}\oplus W_{7,0}$; Time slot indexed by (130): $W_{1,2}\oplus W_{8,1}$;
Time slot indexed by (120): $W_{1,3}\oplus W_{7,1}$;
Time slot indexed by (110): $W_{1,4}\oplus W_{6,1}$;
Time slot indexed by (200): $W_{2,0}\oplus W_{5,2}$;
Time slot indexed by (220): $W_{2,3}\oplus W_{7,2}$;
Time slot indexed by (210): $W_{2,4}\oplus W_{6,2}$;
Time slot indexed by (300): $W_{3,0}\oplus W_{5,3}$;
Time slot indexed by (340): $W_{3,1}\oplus W_{9,3}$;
Time slot indexed by (310): $W_{3,4}\oplus W_{6,3}$;
Time slot indexed by (400): $W_{4,0}\oplus W_{5,4}$;
Time slot indexed by (440): $W_{4,1}\oplus W_{9,4}$;
Time slot indexed by (430): $W_{4,2}\oplus W_{8,4}$.
Then each user can decode its requested file since each file $W_n$ can be recovered by any 4 packets out of $\{W_{n,j}\mid j\in[0, 4]\}$. E.g., user 1 first decodes the required packets $W_{1,2},W_{1,3}$ and $W_{1,4}$ from the coded packets $W_{1,2}\oplus W_{8,1},W_{1,3}\oplus W_{7,1}$ and $W_{1,4}\oplus W_{6,1}$, respectively. Then $W_{1,0}$ can be obtained from $W_{1,4}=W_{1,0}+ W_{1,1}+ W_{1,2}+ W_{1,3}$ since $W_{1,1}$ has been cached by user 1. Hence, array $\overline{\textbf{P}}_2$ can realize a coded caching scheme with a memory ratio of $\frac{M}{N}=\frac{1}{4}$, a transmission rate of $R=\frac{15}{4}$ and a subpacketization level of $F=4$, which is in coincidence with {\it Theorem 3}.
%\begin{table}[H]
%	\centering  % ÏÔʟλÖÃΪÖÐŒä
%\setlength{\abovecaptionskip}{0cm}
%\setlength{\belowcaptionskip}{0.2cm}
%	\caption{Delivery Steps in (1)}  % ±ížñ±êÌâ
%	\label{}  % ÓÃÓÚË÷Òý±ížñµÄ±êÇ©
	%×ÖÄžµÄžöÊý¶ÔÓŠÁÐÊý£¬|Žú±í·ÖžîÏß
	% lŽú±í×ó¶ÔÆ룬cŽú±íŸÓÖУ¬rŽú±íÓÒ¶ÔÆë
%	\begin{tabular}{|c|c|c|c|c|c|}
%		\hline  % ±ížñµÄºáÏß
		 %¿ÉÒÔ±ÜÃâÎÄ×ÖÆ«ÉÏÀŽµ÷ÕûÎÄ×ÖÓëÉϱߜçµÄŸàÀë
%		Time Slot&Transmitted Signal \\  % ±ížñÖеÄÄÚÈÝ£¬ÓÃ&·Ö¿ª£¬\\±íÊŸÏÂÒ»ÐÐ
%		\hline
		 %¿ÉÒÔ±ÜÃâÎÄ×ÖÆ«ÉÏ
%		0&$W_{0,1}\oplus W_{9,0}$\\\hline
%1&$W_{0,2}\oplus W_{8,0}$ \\\hline
%2&$W_{0,3}\oplus W_{7,0}$\\\hline
%3&$W_{1,2}\oplus W_{8,1}$\\\hline

%4&$W_{1,3}\oplus W_{7,1}$\\\hline
%5&$W_{1,4}\oplus W_{6,1}$ \\\hline
%6&$W_{2,0}\oplus W_{5,2}$\\\hline
%7&$W_{2,3}\oplus W_{7,2}$\\\hline

%8&$W_{2,4}\oplus W_{6,2}$\\\hline
%9&$W_{3,0}\oplus W_{5,3}$ \\\hline
%10&$W_{3,1}\oplus W_{9,3}$\\\hline
%11&$W_{3,4}\oplus W_{6,3}$\\\hline

%12&$W_{4,0}\oplus W_{5,4}$ \\\hline
%13&$W_{4,1}\oplus W_{9,4}$\\\hline
%14&$W_{4,2}\oplus W_{8,4}$\\\hline

%
%		\hline
%	\end{tabular}
%\end{table}
Based on {\it Lemma 7}, the scheme in {\it Theorem 2} can also be further improved with a smaller memory ratio and subpacketization level. As the proof of the following result is similar to that of {\it Theorem 3}, it is omitted.

\emph{\textbf{Theorem 4}}. Given any $q, m, t\in\mathbb{N}^{+}$ with
$q\geq 2$ and $t<m$, let $z_{r}^{\ast}$ denote the minimal integer in the set $\mathcal{G}_r=\{z\mid \lfloor\frac{q-1}{q-z}\rfloor=r, z\in[1,q-1]\}$, where $r\in[1,q-1]$.
There exists an $([{m-1\choose t}\lfloor\frac{q-1}{q-z}\rfloor^{t}+{m\choose t}-{m-1\choose t}]q^{t}, M, N)$ coded
caching scheme with a memory ratio of $\frac{M}{N}=\frac{1-(\frac{q-z_{r}^{\ast}}{q})^{t}}{1-(\frac{q-z_{r}^{\ast}}{q})^{t}+(\frac{q-z}{q})^{t}}$, a transmission rate of $R=\frac{(q-z)^{t}}{1-(\frac{q-z_{r}^{\ast}}{q})^{t}+(\frac{q-z}{q})^{t}}$, and a subpacketization level of $F= q^{m-1}[1+(\frac{q-z}{q})^{t}-(\frac{q-z_{r}^{\ast}}{q})^{t}]$.
\section{Analyses of the New PDA Schemes}
This section further analyzes the proposed PDA schemes, consolidating the results given in Table II and establishing a comparison with the existing schemes in Table I.
%we give comparisons of some key parameters among the proposed schemes in Table II and some existing ones in Table I.
\subsection{Comparison of the Schemes in Theorems 1, 3, \cite{19} and \cite{29}}
Some numerical evaluations will be presented to compare the proposed schemes in {\it Theorems 1} and {\it 3} with the existing
one in \cite{19}. Let us assume $m=3,t=2,q=9$ and $z\in[1,8]$. Fig.3 shows that the proposed scheme in {\it Theorem 1} has a smaller subpacketization level than the scheme in \cite{19} with the same number of users, memory ratio and transmission rate. Moreover, the subpacketization level of the scheme in {\it Theorem 3} is smaller than the schemes in \cite{19} and {\it Theorem 1}, while maintaining approximately the same transmission rate.
%\begin{figure}[h]
%\begin{center}
%\includegraphics[scale=0.41]{2.png}
%\caption{\label{1}The packet number of schemes from Theorem 1 and \cite{19} when $K=243$.}
% \end{center}
% \end{figure}
\begin{figure}[H]
%\centering
\subfigure[]{
\begin{minipage}{8.5cm}
%\centering
\includegraphics[scale=0.47]{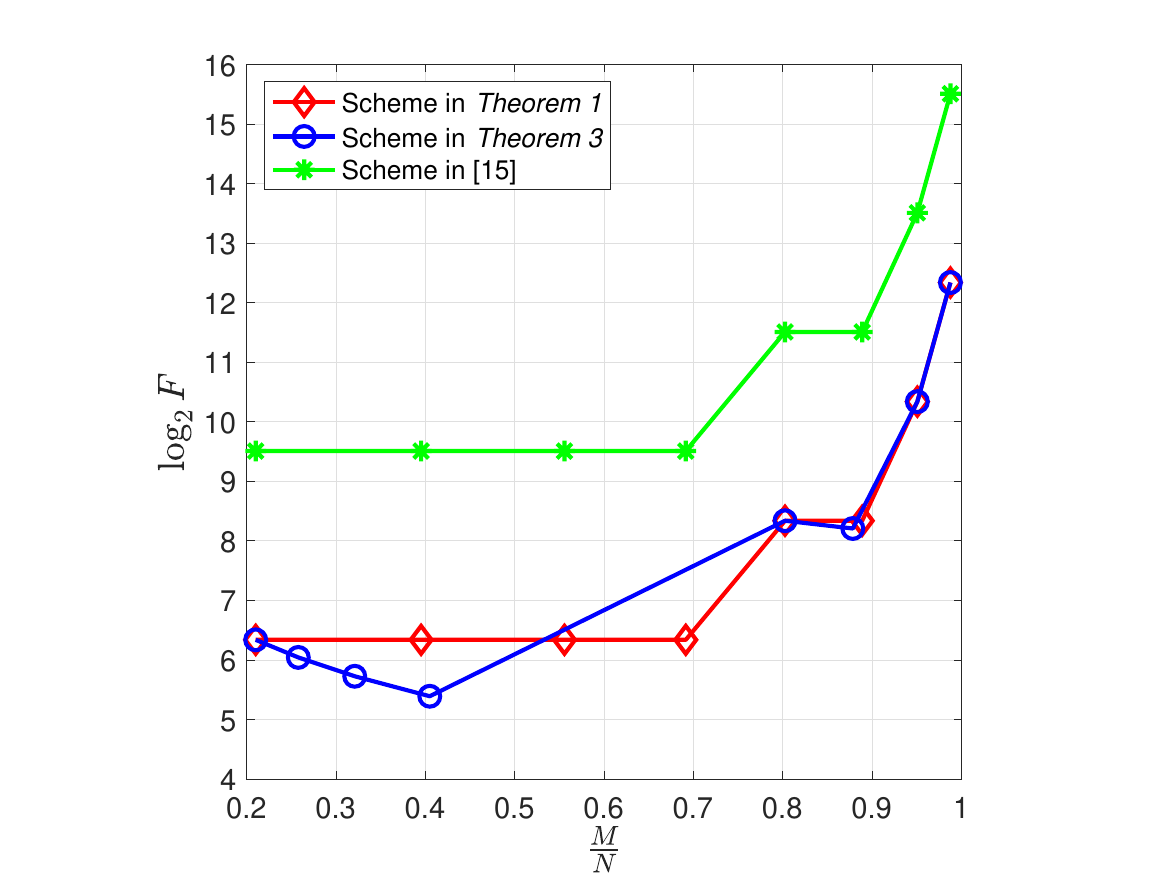}
\end{minipage}%
}%
\subfigure[]{
\begin{minipage}{8.5cm}
%\centering
\includegraphics[scale=0.47]{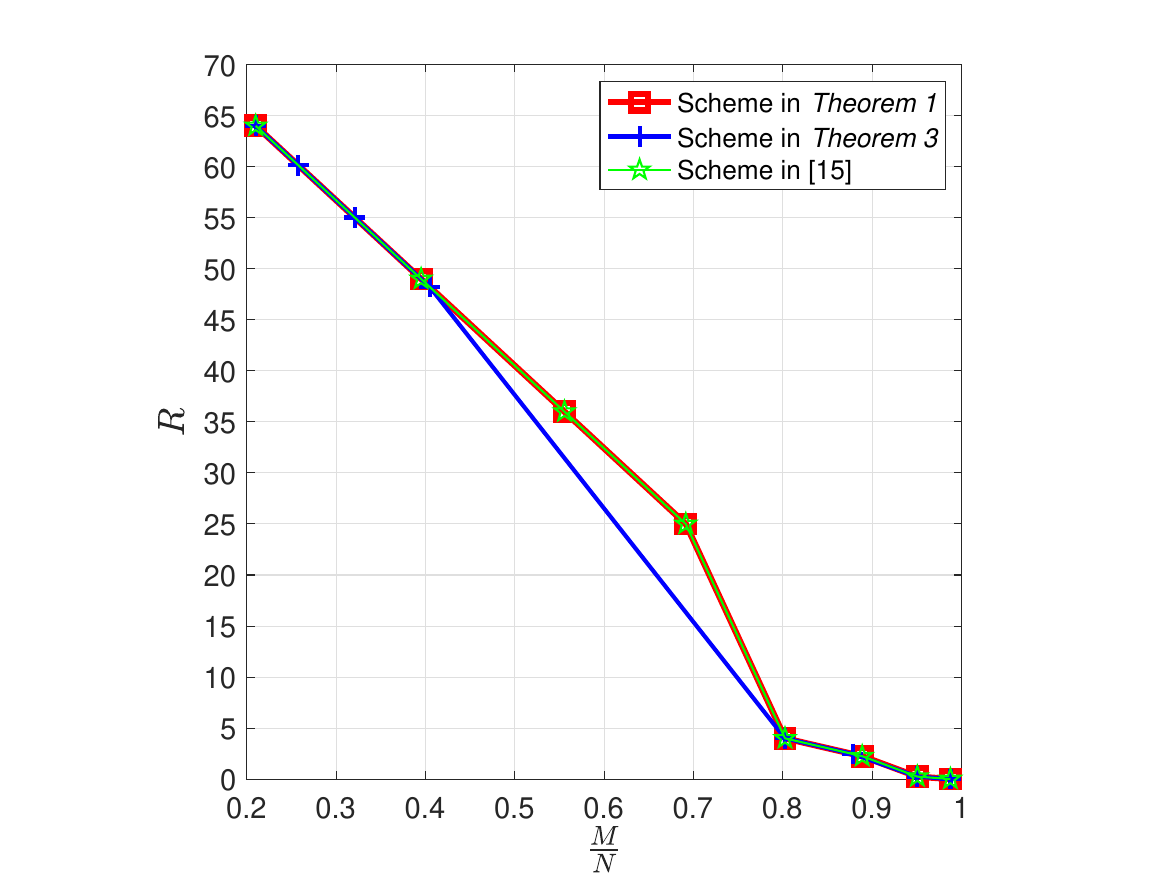}
\end{minipage}
}
\caption{(a) The subpacketization level of schemes in {\it Theorems 1, 3} and \cite{19} when $K=243$; (b) The transmission rate of schemes in {\it Theorems 1, 3} and \cite{19} when $K=243$.}
\end{figure}
When $z = 1$, our scheme in {\it Theorem 1} is the same as the scheme in \cite{29}. However, our scheme generalizes the scheme
in \cite{29} with a more flexible memory size.
\subsection{Comparison of the Schemes in Theorem 2 and \cite{21,16}}
We first discuss the performance of our scheme in {\it Theorem 2} by comparing it with the one in \cite{21}.
%Unfortunately, the theoretical analyses is very complex and cannot yield much intuition. Instead, we illustrate the advantages of our proposed scheme by numerical comparisons in Table III.
The advantage of our proposed scheme will be illustrated in Table III.
For convenience, the parameters of the schemes in {\it Theorem 2} and \cite{21} are written as $(q,z,m,t)$ and $(q,m,t,k)$, respectively. It can be seen that our new scheme in {\it Theorem 2} has a smaller subpacketization level, lower or approximately the same memory ratio and a smaller transmission rate, and is able to serve more users simultaneously.
\begin{center}
\includegraphics[scale=0.237]{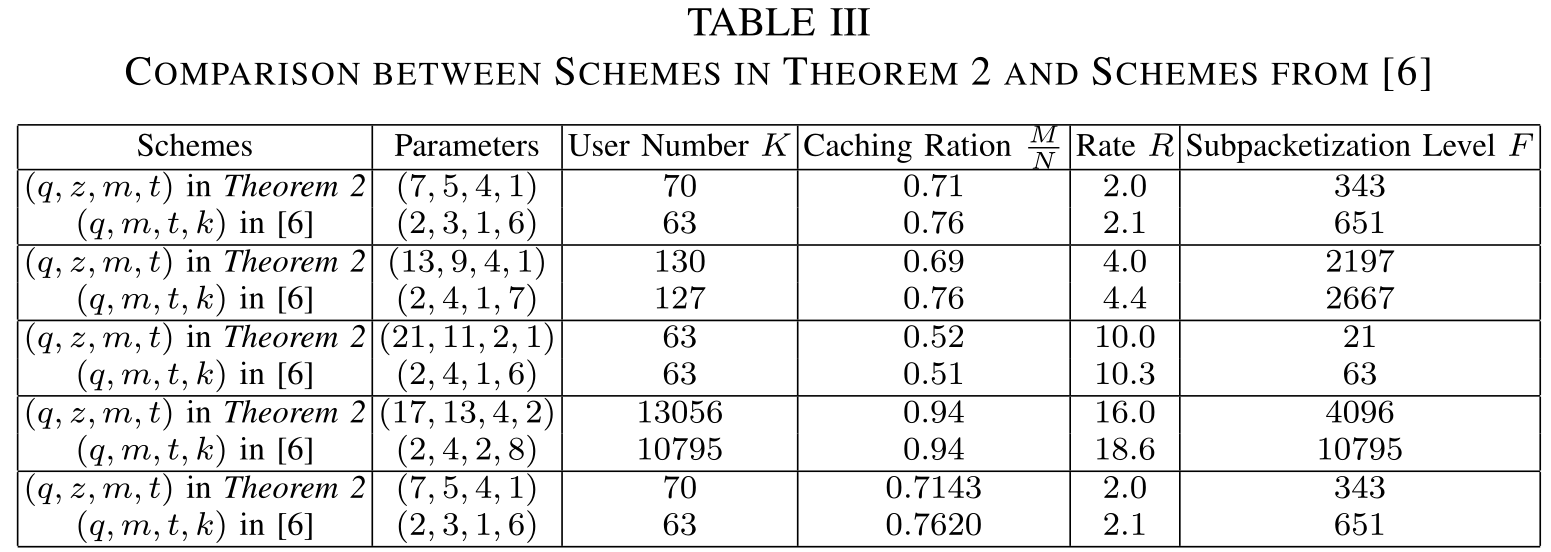}
\end{center}
%\begin{table}[H]
%\center
%\caption{Comparison between Schemes in Theorem 2 and Schemes from \cite{21}}
%\label{com1}
 %   \renewcommand\arraystretch{1}
 %  \setlength{\tabcolsep}{0.5mm}{
%\begin{tabular}{|c|c|c|c|c|c|}
%\hline
%Schemes & Parameters & User number $K$  &Caching ration $\frac{M}{N}$ & Rate $R$ &Subpacketization level $F$\\
%\hline
%$(q,z,m,t)$ in Theorem 2 & $(7,5,4,1)$ & $70$  & $0.71$ & $2.0$& $343$\\
%$(q,m,t,k)$ in \cite{21}  & $(2,3,1,6)$ & $63$  & $0.76$ & $2.1$ & $651$\\
%\hline
%$(q,z,m,t)$ in Theorem 2 & $(13,9,4,1)$ & $130$  & $0.69$ & $4.0$& $2197$\\
%$(q,m,t,k)$ in \cite{21}  & $(2,4,1,7)$ & $127$  & $0.76$ & $4.4$ & $2667$\\
%\hline
%$(q,z,m,t)$ in Theorem 2 & $(21,11,2,1)$ & $63$  & $0.52$ & $10.0$& $21$\\
%$(q,m,t,k)$ in \cite{21}  & $(2,4,1,6)$ & $63$  & $0.51$ & $10.3$ & $63$\\
%\hline
%$(q,z,m,t)$ in Theorem 2 & $(17,13,4,2)$ & $13056$  & $0.94$ & $16.0$& $4096$\\
%$(q,m,t,k)$ in \cite{21}  & $(2,4,2,8)$ & $10795$  & $0.94$ & $18.6$& $10795$ \\
%\hline
%$(q,z,m,t)$ in Theorem 2 & $(7,5,4,1)$ & $70$  & $0.7143$ & $2.0$& $343$\\
%$(q,m,t,k)$ in \cite{21}  & $(2,3,1,6)$ & $63$  & $0.7620$ & $2.1$ & $651$\\
%\hline
%\end{tabular}}
%\end{table}

 We now consider the comparison between the scheme in {\it Theorem 2} and the one in \cite{16}. Let $z=q-1$ in {\it Theorem 2}. Table IV shows that the scheme in {\it Theorem 2} has advantage in the number of users and subpacketization level at the cost of some transmission rate. Furthermore,
it is interesting that when $t$ approximates to $\frac{m}{2}$, we have $K=[{m-1\choose t}(q-1)^{t}+{m\choose t}-{m-1\choose t}]q^{t}$,
and it is near to $q^{m-1}$, i.e., the subpacketization level $F$ is linear or polynomial to the number of users $K$.
\begin{center}
\includegraphics[scale=0.223]{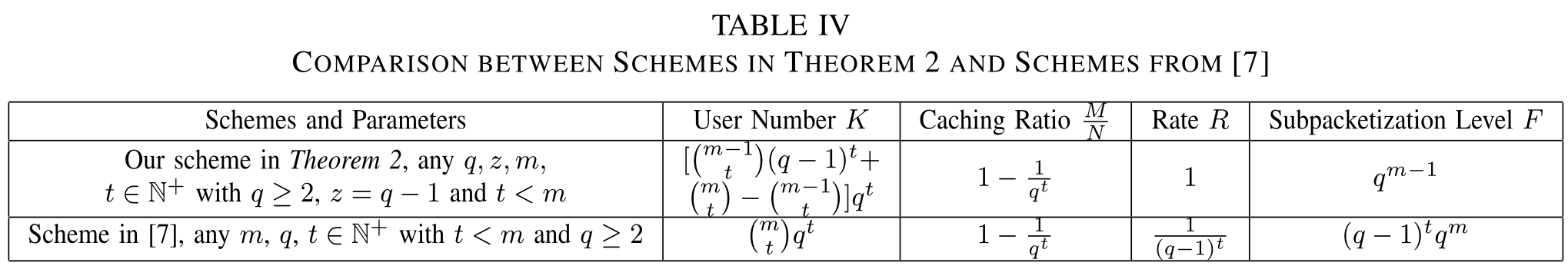}
\end{center}
%{\begin{table}[H]
%\center
%\caption{Comparison between Schemes in Theorem 2 and Schemes from \cite{16}}
%\resizebox{\textwidth}{10mm}{
%\setlength{\arraycolsep}{0.25pt}
%\begin{tabular}{|c|c|c|c|c|c|}
%\hline
%Schemes and parameters &User number $K$  &Caching ratio $\frac{M}{N}$   &Rate $R$   &Subpacketization level $F$    \\ \hline
%\tabincell{c}{Our scheme in Theorem 2, any $q, z, m,$\\ $t\in\mathbb{N}^{+}$ with
%$q\geq2$, $z=q-1$ and $t<m$} &\tabincell{c}{$[{m-1\choose t}(q-1)^{t}+$\\${m\choose t}-{m-1\choose t}]q^{t}$} &$1-\frac{1}{q^{t}}$ &$1$& %$q^{m-1}$\\\hline
%\tabincell{l}{Scheme in \cite{16}, any $m$, $q$, $t\in  \mathbb{N}^+$ with $t<m$ and $q\geq2$}
%&${m\choose t}q^{t}$ &$1-\frac{1}{q^{t}}$&$\frac{1}{(q-1)^{t}}$& $(q-1)^{t}q^{m}$\\   \hline
%\end{tabular} }
%\end{table}}
\subsection{Comparison of the Schemes in Theorem 4 and \cite{22}}
We further compare the performance of our scheme in {\it Theorem 4} and the one in \cite{22}. For convenience, the parameters of the schemes in {\it Theorem 4} and \cite{22} are written as $(q,z,m,t,z_{r}^{\ast})$ and $(m,k,t,q)$, respectively. Table V shows that our new scheme in {\it Theorem 4} has a smaller subpacketization level, lower or approximately the same memory ratio and a larger number of users, while sacrifices some transmission rate.
\begin{center}
\includegraphics[scale=0.242]{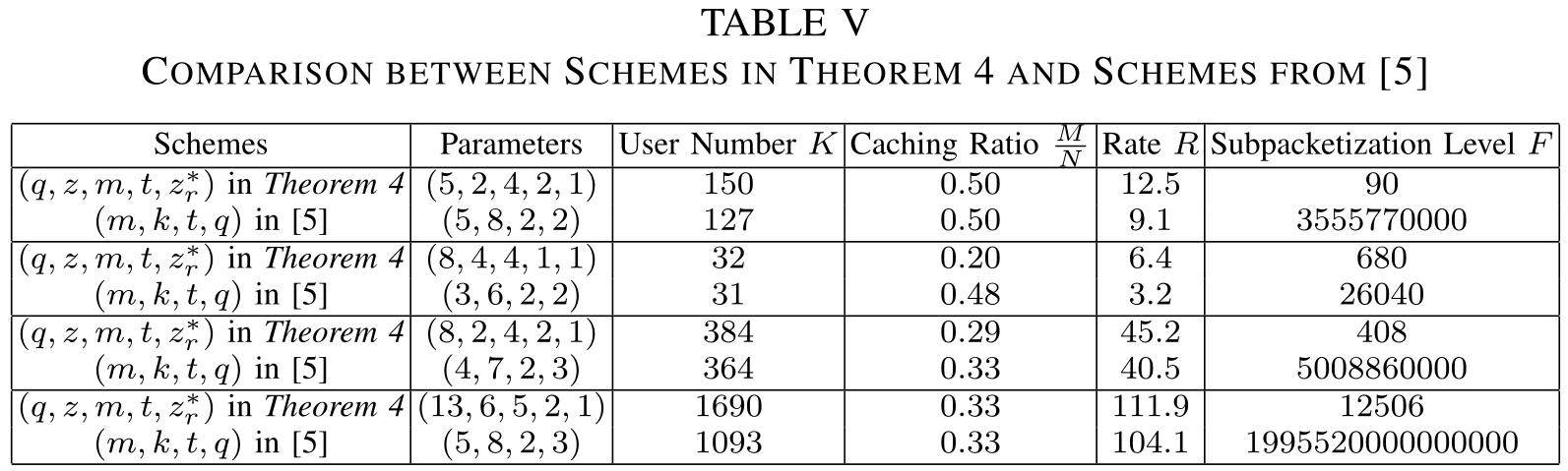}
\end{center}
%\begin{table}[H]
%\center
%\caption{Comparison between Schemes in Theorem 4 and Schemes from \cite{22}}
%\label{com1}
%   \renewcommand\arraystretch{1}
%    \setlength{\tabcolsep}{0.5mm}{
%\begin{tabular}{|c|c|c|c|c|c|}
%\hline
% Schemes & Parameters &User number $K$  &Caching ratio $\frac{M}{N}$ &Rate $R$&Subpacketization level $F$ \\
%\hline
%$(q,z,m,t,z_{r}^{\ast})$ in Theorem 4 & $(5,2,4,2,1)$ & $150$  & $0.50$ & $12.5$& $90$\\
%$(m,k,t,q)$ in \cite{22}  & $(5,8,2,2)$ & $127$  & $0.50$ & $9.1$& $3555770000$ \\
%\hline
%$(q,z,m,t,z_{r}^{\ast})$ in Theorem 4 & $(8,4,4,1,1)$ & $32$  & $0.20$ & $6.4$& $680$\\
%$(m,k,t,q)$ in \cite{22}  & $(3,6,2,2)$ & $31$  & $0.48$ & $3.2$& $26040$ \\
%\hline
%$(q,z,m,t,z_{r}^{\ast})$ in Theorem 4 & $(8,2,4,2,1)$ & $384$  & $0.29$ & $45.2$& $408$\\
%$(m,k,t,q)$ in \cite{22}  & $(4,7,2,3)$ & $364$  & $0.33$ & $40.5$ & $5008860000$\\
%\hline
%$(q,z,m,t,z_{r}^{\ast})$ in Theorem 4 & $(13,6,5,2,1)$ & $1690$  & $0.33$ & $111.9$& $12506$\\
%$(m,k,t,q)$ in \cite{22}  & $(5,8,2,3)$ & $1093$  & $0.33$ & $104.1$ & $1995520000000000$\\
%\hline
%\end{tabular}}
%\end{table}
\section{Conclusion}
 This paper has proposed a novel construction of PDA via POAs. Based on the construction, some coded caching schemes have been obtained with a low subpacketization level and a more flexible memory size. The first PDA scheme achieves an improved subpacketization level with the same number of users, memory ratio and transmission rate over the existing one. The second PDA scheme enables a larger number of users and a smaller subpacketization level than the first scheme when the memory ratio is greater than $\frac{1}{2}$. Moreover, two new coded caching schemes with the coded placement based on the proposed PDAs have also been considered. Our numerical analyses have indicated that the proposed schemes have better performance than the existing ones in literature.
\appendices

%\emph{\textbf{Remark} 2}. It can be seen that the PDA scheme in \cite{19} has a more flexible memory size than the scheme in \cite{16}, while the row indices of the PDA of \cite{19} are obtained by just reusing $\lfloor\frac{q-1}{q-z}\rfloor^{t}$ times of the row indices of the PDA of \cite{16} and the entry rule of the PDA of \cite{19} can be regarded as a generalization of the PDA of \cite{16}. Moreover, the PDA scheme in \cite{29} has a lower subpacketization than that of the scheme in \cite{16} with the same number of users, memory ratio and transmission rate.  Unfortunately it is hard to obtain a similar result as that of [15] but smaller subpacketization by simply changing the entry rule or reusing the same OA$(m,q,m-1)$ which is used in the construction of \cite{29}, i.e., it is hard to use the similar method in the construction of \cite{19} based on the construction of \cite{29}. Furthermore, the entry rule defined in the construction of \cite{29} is different from that of the PDA generated in the construction of \cite{19}. This implies that it is impossible to obtain our PDAs by just deleting some rows of the PDA generated by the construction of \cite{19}. So in order to obtain the result of this paper, we use the different method in the construction of \cite{19}, i.e., by means of using $\lfloor\frac{q-1}{q-z}\rfloor^{t}$ times of the proper OA$(m,q,m-1)$.

%\begin{figure}[h]
%\begin{center}
%\includegraphics[scale=0.41]{2.png}
%\caption{\label{1}The packet number of schemes from Theorem 1 and \cite{19} when $K=243$.}
% \end{center}
% \end{figure}

\section{Proof of Theorem 1}
\label{sec:proof of theorem-1}
In order to prove {\it Theorem 1}, we need the following results.

\emph{\textbf{Proposition 1}}. Array \textbf{P} generated from {\it Construction 1} satisfies condition C3 of {\it Definition 1}.
%\label{propo-1} Let \textbf{P}
% denote an array generated from Construction 1. If two entries are the same,
% i.e., $p_{{(\textbf{f}^{(j)}_s, \textbf{g}_j),(\mathcal{I}, \textbf{c})}}=p_{{(\textbf{f}^{(j')}_{s'}, \textbf{g}_{j'}),(\mathcal{I}', \textbf{c}')}}=(\textbf{v}, o(\textbf{v}))$, the following two properties hold:
%(a). Vector entry $(\textbf{v}, o(\textbf{v}))$ occurs in different rows and different columns.
%(b). The subarray formed by different rows $(\textbf{f}^{(j)}_s, \textbf{g}_j)$, $(\textbf{f}^{(j')}_{s'}, \textbf{g}_{j'})$ and different columns $(\mathcal{I}, \textbf{c})$, $(\mathcal{I}', \textbf{c}')$ must be in one of the following forms
%\begin{equation*}
%\left(
%  \begin{array}{cc}
%    (\textbf{v}, o(\textbf{v})) & \ast \\
%    \ast & (\textbf{v}, o(\textbf{v})) \\
%  \end{array}
%\right)\text{,} \left(
%  \begin{array}{cc}
%    \ast & (\textbf{v}, o(\textbf{v})) \\
%    (\textbf{v}, o(\textbf{v})) & \ast \\
%  \end{array}
%\right).
%\end{equation*}
\begin{proof}
 Suppose that $p_{{(\textbf{f}^{(j)}_{s}, \textbf{g}_{j}),(\mathcal{I}, \textbf{c})}}=p_{{(\textbf{f}^{(j')}_{s'}, \textbf{g}_{j'}),(\mathcal{I}', \textbf{c}')}}=(\textbf{v},o(\textbf{v}))$, where
$$\textbf{f}^{(j)}_{s}=(f^{(j)}_{s,0},f^{(j)}_{s,1},\cdots, f^{(j)}_{s,m-1}); \textbf{f}^{(j')}_{s'}=(f^{(j')}_{s',0},f^{(j')}_{s',1},\cdots, f^{(j')}_{s',m-1});\mathcal{I}=(\xi_0, \xi_1,\cdots, \xi_{t-1});$$
$$\mathcal{I}'=(\xi'_0, \xi'_1,\cdots, \xi'_{t-1});\textbf{g}_j=(g^{(j)}_0, g^{(j)}_1,\cdots, g^{(j)}_{t-1}); \textbf{g}_{j'}=(g^{(j')}_0, g^{(j')}_1,\cdots, g^{(j')}_{t-1});$$
$$\textbf{c}=(c_0, c_1,\cdots, c_{t-1}); \textbf{c}'=(c'_0, c'_1,\cdots, c'_{t-1});\textbf{v}=(v_0, v_1, \cdots, v_{m-1}).$$
In order to prove C3-(a), i.e., the vector entry $(\textbf{v}, o(\textbf{v}))$ occurs in different rows and different columns, it is sufficient to prove that it is impossible for vector \textbf{v} to appear more than once in some row.
%there is no vector \textbf{v} occurring more than once in some row.
Assume that vector $\textbf{v}$ appears more than once in some row, then let us consider the following two cases under the condition $(\textbf{f}^{(j)}_s, \textbf{g}_j)=(\textbf{f}^{(j')}_{s'}, \textbf{g}_{j'})$, i.e., $s=s'$ and $j=j'$.

\textbf{Case 1}: $\mathcal{I}=\mathcal{I}'$. If $\textbf{c}\neq \textbf{c}'$, without loss of generality, let us assume that $c_0 \neq c'_0$. It follows from {\it Construction 1} that $v_{\xi_0}=c_0-g^{(j)}_0(q-z)=c'_0-g^{(j)}_0(q-z)$, i.e., $c_0=c'_0$, which contradicts our hypothesis. This implies that
$\textbf{c}= \textbf{c}'$, i.e., $(\mathcal{I}, \textbf{c})= (\mathcal{I}', \textbf{c}')$.

\textbf{Case 2}: $\mathcal{I}\neq\mathcal{I}'$. There must exist
two distinct integers, say $\alpha$, $\alpha'\in[0, m-1]$, satisfying $\alpha\in\mathcal{I}, \alpha\notin\mathcal{I}'$ and $\alpha'\in\mathcal{I}', \alpha'\notin\mathcal{I}$.
Without loss of generality, let us assume that $\alpha=\xi_0, \alpha'=\xi'_0$. Based on {\it Construction 1}, we have $v_{\xi_0}=c_0-g^{(j)}_0(q-z)=f^{(j)}_{s,\xi_0}$
and $v_{\xi'_0}=c'_0-g^{(j)}_0(q-z)=f^{(j)}_{s,\xi'_0}$. This implies that $p_{{(\textbf{f}^{(j)}_s, \textbf{g}_j),(\mathcal{I}, \textbf{c})}}=\ast$ since $f^{(j)}_{s,\xi_0}=c_0-g^{(j)}_0(q-z)\in\{c_0, c_0-1, \cdots, c_0-(z-1)\}$, which contradicts $p_{{(\textbf{f}^{(j)}_s, \textbf{g}_j),(\mathcal{I}, \textbf{c})}}=(\textbf{v}, o(\textbf{v}))$.

%Similarly, we can also show that $p_{{(\textbf{f}^{(j)}_s, \textbf{g}_j),(\mathcal{I}', \textbf{c}')}}=\ast$.
 Based on the above argument, it can be seen that C3-(a) of {\it Definition 1} holds. Next we show that C3-(b) of {\it Definition 1} also holds.

 Suppose that vector $\textbf{v}$ occurs in distinct rows and columns, i.e., rows $(\textbf{f}^{(j)}_s, \textbf{g}_j)$, $(\textbf{f}^{(j')}_{s'}, \textbf{g}_{j'})$ and columns $(\mathcal{I}, \textbf{c}), (\mathcal{I}', \textbf{c}')$. If $\mathcal{I}\neq\mathcal{I}'$, with the similar argument of Case 2, it can be seen that C3-(b) of {\it Definition 1} holds.
It remains to show that C3-(b) of {\it Definition 1} holds for $\mathcal{I}=\mathcal{I}'$.  To do this we just need to consider $\textbf{g}_j\neq\textbf{g}_{j'}$, since it is impossible for vector \textbf{v} to appear in both entries $p_{{(\textbf{f}^{(j)}_{s}, \textbf{g}_{j}),(\mathcal{I}, \textbf{c})}}$ and $p_{{(\textbf{f}^{(j')}_{s'}, \textbf{g}_{j}),(\mathcal{I}, \textbf{c}')}}$. Let us assume that $p_{{(\textbf{f}^{(j)}_s, \textbf{g}_j),(\mathcal{I}, \textbf{c}')}}\neq\ast$, then we have $f^{(j)}_{s,\xi_i}\in\{c_i'+1, c_i'+2, \cdots, c_i'+(q-z)\}$ for any $i\in[0,t-1]$, i.e., there exists an integer $\gamma\in[1, q-z]$ satisfying $f^{(j)}_{s,\xi_i}=c_i'+\gamma$. Based on {\it Construction 1}, we have $c_i'=c_i+(q-z)(g^{(j')}_i-g^{(j)}_i)$. Hence, we obtain the following equation
\begin{equation}
f^{(j)}_{s,\xi_i}-c_i=(q-z)(g^{(j')}_i-g^{(j)}_i)+\gamma.
\end{equation}
This is impossible for $g^{(j)}_i<g^{(j')}_i$ due to $q-z+1<(q-z)(g^{(j')}_i-g^{(j)}_i)+\gamma<q$ and $0< f^{(j)}_{s,\xi_i}-c_i<q-z+1$ by the fact
$g^{(j)}_i,g^{(j')}_i\in [0, \lfloor\frac{q-1}{q-z}\rfloor-1]$ and $q<2z$. If $g^{(j)}_i>g^{(j')}_i$, the equation of (14) can be written as $f^{(j)}_{s,\xi_i}-c_i+(q-z)(g^{(j)}_i-g^{(j')}_i)=\gamma$, which is also impossible since $q-z<f^{(j)}_{s,\xi_i}-c_i+(q-z)(g^{(j)}_i-g^{(j')}_i)<q$ and $1\leq\gamma\leq q-z$.
 Therefore, we have $p_{{(\textbf{f}^{(j)}_s, \textbf{g}_j),(\mathcal{I}, \textbf{c}')}}=\ast$. Similarly, we can also show that $p_{{(\textbf{f}^{(j')}_s, \textbf{g}_{j'}),(\mathcal{I}, \textbf{c})}}=\ast$. So C3-(b) of {\it Definition 1} holds.
\end{proof}
%Proposition 1 shows that the array \textbf{P} generated from Construction 1 satisfies C3 of Definition 1. In order to prove that array \textbf{P}
%is a PDA, we just need to prove that the number of $``\ast"$s in each column of \textbf{P} is a constant.
\emph{\textbf{Proposition 2}}. Given $q, m, z, t\in\mathbb{N}^{+}$ with $q\geq 2, z<q$ and $t<m$, let $\textbf{P}=(
     \textbf{P}_{0};
     \cdots;
     \textbf{P}_{\lfloor\frac{q-1}{q-z}\rfloor^{t}-1}
   )$
denote an array generated from {\it Construction 1} with row indices set $\mathcal{F}=\mathop\bigcup_{j\in[0,\lfloor\frac{q-1}{q-z}\rfloor^{t}-1]}\{(\textbf{f}_s^{(j)},\textbf{g}_j)\mid s\in[0,q^{m-1}-1]\}$ and $\textbf{f}_s^{(j)}\in\textbf{F}_{\textbf{g}_j}$.
Then all the columns of $\textbf{P}_j$ have the same number of $``\ast"$s, i.e., \textbf{P} is a PDA. So the memory ratio of the $({m\choose t}q^{t}, M, N)$ caching system realized by \textbf{P} is $\frac{M}{N}=1-(\frac{q-z}{q})^{t}$.
\begin{proof}
 Note that an OA$(m,q,m-1)$ is also an OA$_{q^{m-t-1}}(m,q,t)$. This implies that $\textbf{F}_{\textbf{g}_j}$ defined in {\it Construction 1} is also an OA$_{q^{m-t-1}}(m,q,t)$. Given any column $(\mathcal{I},\textbf{c})=(\{\xi_0,\xi_1,\cdots,\xi_{t-1}\},$\\$(c_0,c_1,\cdots,c_{t-1}))$ of $\textbf{P}_j$, the number of row vectors $\textbf{f}^{(j)}_{s}=(f^{(j)}_{s,0},f^{(j)}_{s,1},\cdots,f^{(j)}_{s,m-1})$ of $\textbf{F}_{\textbf{g}_j}$ such that $f^{(j)}_{s,\xi_i}\notin\{c_i,c_i-1,\cdots,c_i-(z-1)\}$ for $i\in[0,t-1]$
 is $(q-z)^{t}q^{m-t-1}$, since each $t$-length row vector appears $q^{m-t-1}$ times in $\textbf{F}_{\textbf{g}_j}$. Then by the entry rule of {\it Construction 1}, the number of non-star entries in each column of $\textbf{P}_j$ is $(q-z)^{t}q^{m-t-1}$, i.e., the number of $``\ast"$s in each column of $\textbf{P}_j$ is $q^{m-1}-(q-z)^{t}q^{m-t-1}$. So the memory ratio of the $({m\choose t}q^{t}, M, N)$ caching system realized by \textbf{P} is $\frac{M}{N}=1-(\frac{q-z}{q})^{t}$.
 \end{proof}
% The following proposition shows the combinatorial property of the the array generated in Construction 1.
\emph{\textbf{Proposition 3}}.
Given $q, m, z, t\in\mathbb{N}^{+}$ with $q\geq 2, z<q$ and $t<m$, let $\textbf{P}=(
     \textbf{P}_{0};
     \cdots;
     \textbf{P}_{\lfloor\frac{q-1}{q-z}\rfloor^{t}-1}
   )$
%\begin{equation*}
%\textbf{P}=\left(\begin{matrix}
%     \textbf{P}_{0}\\
 %    \vdots\\
 %    \textbf{P}_{j}\\
 %    \vdots\\
 %    \textbf{P}_{\lfloor\frac{q-1}{q-z}\rfloor^{t}-1}
 %  \end{matrix}\right)
 %  \end{equation*}
denote a $q^{m-1}\lfloor\frac{q-1}{q-z}\rfloor^{t}\times {m\choose t}q^{t}$ array generated from {\it Construction 1} with row indices set $\mathcal{F}=\mathop\bigcup_{j\in[0,\lfloor\frac{q-1}{q-z}\rfloor^{t}-1]}\{(\textbf{f}_s^{(j)},\textbf{g}_j)\mid s\in[0,q^{m-1}-1]\}$ and $\textbf{f}_s^{(j)}\in\textbf{F}_{\textbf{g}_j}$. Then each $\textbf{P}_j$ in $\textbf{P}$ satisfies

(a). Each vector \textbf{v} in $\textbf{P}_j$ occurs in exactly $m\choose t$ columns, and each vector \textbf{v} in $\textbf{P}_j$ occurs the same number of times in each column where \textbf{v} occurs;

(b). Each vector \textbf{v} occurring in $\textbf{P}_i$ must occur in $\textbf{P}_j$ for any $i,j\in[0,\lfloor\frac{q-1}{q-z}\rfloor^{t}-1]$.

In order to prove {\it Proposition 3}, the following result is useful.

\emph{\textbf{Proposition 4}} \cite{29}. Given $m, q, z, t\in\mathbb{N}^{+}$ with $t\leq m$ and $z=1$, let $\textbf{P}_{0}$
denote a $q^{m-1}\times {m\choose t}q^{t}$ array obtained from (9) such that $\mathcal{F}=$OA$(m,q,m-1)$. Then each vector \textbf{v} in $\textbf{P}_0$ occurs in exactly $m\choose t$ columns, and each vector \textbf{v} in $\textbf{P}_0$ occurs the same number of times in each column where \textbf{v} occurs.

Now let us prove {\it Proposition 3}.
\begin{proof}
(a). We first consider $\textbf{P}_0$ with row indices generated by $\mathcal{F}_{\textbf{g}_0}^{(0)}=\{(\textbf{f}_s^{(0)},\textbf{g}_0)\mid s\in[0, q^{m-1}-1]\}$, where $\textbf{f}_s^{(0)}\in \textbf{F}_{\textbf{g}_0}$, $\textbf{g}_0=(0,0,\cdots,0)$, and $\sum_{r=0}^{m-1}f_{s,r}^{(0)}=0$ for $s\in[0, q^{m-1}-1]$. Based on {\it Proposition 4}, It can be seen that $\textbf{P}_0$ satisfies the statement of (a) when $z=1$. If $z>1$, $\textbf{P}_0$ can be viewed as replacing some vector entries with $``\ast"$ from the case $z=1$. Therefore, in order to prove that $\textbf{P}_0$ satisfies the statement of (a) for any $z\in[2, q-1]$, we just need to prove such vector entries that contain \textbf{v} are replaced by $``\ast"$s uniformly in the column where they occur.
Given any $\textbf{v}=(v_0,v_1,\cdots,v_{m-1})$ in $\textbf{P}_0$ and fixed column $(\mathcal{I},\textbf{c})=(\{\xi_0,\xi_1,\cdots,\xi_{t-1}\},(c_0,c_1,\cdots,c_{t-1}))$, we denote the collection of rows in which $\textbf{v}$ occurs by

$\mathcal{F}_{\mathcal{I},\textbf{v}}^{(0)}=\{(\textbf{f}_s^{(0)},\textbf{g}_0)=((f_{s,0}^{(0)},f_{s,1}^{(0)},\cdots,f_{s,m-1}^{(0)}),\textbf{g}_0)\mid p_{(\textbf{f}_s^{(0)},\textbf{g}_0),(\mathcal{I},\textbf{c})}=(\textbf{v},o(\textbf{v})), s\in[0, q^{m-1}-1]\}.$\\
Similarly, let

$\mathcal{F}^{(0)}_{\mathcal{I}',\textbf{v}}=\{(\textbf{f}_{s'}^{(0)},\textbf{g}_0)=((f_{s',0}^{(0)},f_{s',1}^{(0)},\cdots,f_{s',m-1}^{(0)}),\textbf{g}_0)\mid p_{(\textbf{f}_{s'}^{(0)},\textbf{g}_0),(\mathcal{I}',\textbf{c}')}=(\textbf{v},o(\textbf{v})), s'\in[0, q^{m-1}-1]\}$\\
denote the collection of rows in which $\textbf{v}$ occurs corresponding to column $({\mathcal{I}',\textbf{c}'})=(\{\xi'_0,\xi'_1\cdots,$\\$\xi'_t\},(c'_0,c'_1,\cdots,c'_{t-1}))$, where $|\mathcal{I}\cap\mathcal{I}'|=t-1$.
Note that when $z=1$, there exists a one to one mapping from $\mathcal{F}^{(0)}_{\mathcal{I},\textbf{v}}$ to $\mathcal{F}^{(0)}_{\mathcal{I}',\textbf{v}}$:
$$\psi((\textbf{f}_s^{(0)},\textbf{g}_0))=(\textbf{f}_{s'}^{(0)},\textbf{g}_0), \;\;\;\; (\textbf{f}_s^{(0)},\textbf{g}_0)\in\mathcal{F}_{\mathcal{I},\textbf{v}}^{(0)},\; (\textbf{f}_{s'}^{(0)},\textbf{g}_0)\in\mathcal{F}_{\mathcal{I'},\textbf{v}}^{(0)},$$
where $f_{s,i}^{(0)}=f_{s',i}^{(0)}$ for $i\in(\mathcal{I}\cap\mathcal{I}')\cup([0,m-1]\backslash(\mathcal{I}\cup\mathcal{I}'))$. Assume that vector $\textbf{v}$ appears in row $(\textbf{f}_s^{(0)},\textbf{g}_0)=((f_{s,0}^{(0)},f_{s,1}^{(0)},\cdots,f_{s,m-1}^{(0)}),(0,0,\cdots,0))$ and column $({\mathcal{I},\textbf{c}})$, then vector $\textbf{v}$ must appear in column $({\mathcal{I}',\textbf{c}'})$ and a unique row $(\textbf{f}_{s'}^{(0)},\textbf{g}_0)=((f_{s',0}^{(0)},f_{s',1}^{(0)},\cdots,f_{s',m-1}^{(0)}),(0,0,\cdots,0))$ since $\psi$ is a one to one mapping. Furthermore, assume that there exist two different integers $\xi_w$ and $\xi'_{w'}$ such that $\mathcal{I}\backslash\mathcal{I}'=\{\xi_w\}$ and $\mathcal{I}'\backslash\mathcal{I}=\{\xi'_{w'}\}$. Based on {\it Construction 1}, there exist two integers $\beta, \beta'$ such that
$f_{s,\xi_w}^{(0)}=c_{w}+\beta, f_{s,\xi'_{w'}}^{(0)}=c'_{w'}$,
$f_{s',\xi_w}^{(0)}=c_{w}$, and $f_{s',\xi'_{w'}}^{(0)}=c'_{w'}+\beta'.$
Then we obtain
$c_{w}+\beta+c'_{w'}=f_{s,\xi_w}^{(0)}+f_{s,\xi'_{w'}}^{(0)}=f_{s',\xi_w}^{(0)}+f_{s',\xi'_{w'}}^{(0)}=c'_{w'}+\beta'+c_{w}$ due to $\sum_{r=0}^{m-1}f_{s,r}^{(0)}=0$ for $s\in[0,q^{m-1}-1]$, i.e., $\beta=\beta'$ holds.

Now let us consider $z>1$ for $\textbf{P}_0$. Without loss of generality, suppose that a vector entry that contains $\textbf{v}$ is replaced by $``\ast"$ in row $(\textbf{f}_s^{(0)},\textbf{g}_0)$ and column $({\mathcal{I},\textbf{c}})$, then there must exist a vector entry that contains \textbf{v} converting into $``\ast"$ in row $(\textbf{f}_{s'}^{(0)},\textbf{g}_0)$ and column $({\mathcal{I}',\textbf{c}'})$ since $\beta'=\beta\notin [1,q-z]$. This implies that vector entries that contain \textbf{v }are replaced by $``\ast"$s uniformly in the column where they occur, i.e., array $\textbf{P}_0$ satisfies the statement of (a). In the following we show that there always exists array $\textbf{P}_j$ satisfying the statement of (a) for $j\in[1,\lfloor\frac{q-1}{q-z}\rfloor^{t}-1]$.

When $q<2z$, $1<z$ and $j\in[1,\lfloor\frac{q-1}{q-z}\rfloor^{t}-1]$, let us consider $\textbf{P}_j$ with row indices generated by $\mathcal{F}_{\textbf{g}_j}^{(j)}=\{(\textbf{f}_s^{(j)},\textbf{g}_j)\mid s\in[0, q^{m-1}-1]\}$, where $\textbf{f}_s^{(j)}\in \textbf{F}_{\textbf{g}_j}$ and $\textbf{g}_j=(g^{(j)}_0, g^{(j)}_1, \cdots, g^{(j)}_{t-1})$.
For any $\textbf{v}$ in $\textbf{P}_0$ and fixed column $(\mathcal{I},\textbf{c})=(\{\xi_0,\xi_1,\cdots,\xi_{t-1}\},(c_0,c_1,\cdots,c_{t-1}))$, we also denote the collection of rows in which $\textbf{v}$ occurs by

$\mathcal{F}_{\mathcal{I},\textbf{v}}^{(0)}=\{(\textbf{f}_s^{(0)},\textbf{g}_0)=((f_{s,0}^{(0)},f_{s,1}^{(0)},\cdots,f_{s,m-1}^{(0)}),\textbf{g}_0)\mid p_{(\textbf{f}_s^{(0)},\textbf{g}_0),(\mathcal{I},\textbf{c})}=(\textbf{v},o(\textbf{v})), s\in[0, q^{m-1}-1]\}.$\\
Similarly, suppose that such vector \textbf{v} appears in $\textbf{P}_j$, then the collection of rows in which $\textbf{v}$ occurs in column $(\mathcal{I},\textbf{c}')=(\{\xi_0,\xi_1,\cdots,\xi_{t-1}\}, (c'_0,c'_1,\cdots,c'_{t-1}))$ can be written as\\
$\mathcal{F}_{\mathcal{I},\textbf{v}}^{(j)}=\{(\textbf{f}_{s'}^{(j)},\textbf{g}_j)=((f_{s',0}^{(j)},f_{s',1}^{(j)},\cdots,f_{s',m-1}^{(j)}),\textbf{g}_j)\mid
p_{(\textbf{f}_{s'}^{(j)},\textbf{g}_j),(\mathcal{I},\textbf{c}')}=(\textbf{v},o(\textbf{v})), s'\in[0, q^{m-1}-1]\}.$\\
Since $\textbf{F}_{\textbf{g}_j}$ is a POA$(m, q, m-1)$ such that $\sum_{r=0}^{m-1}f_{s'r}^{(j)}=\sum_{i=0}^{t-1}g^{(j)}_i(q-z)$ for $s'\in[0, q^{m-1}-1]$, we can define a mapping $\phi$ from $\mathcal{F}_{\mathcal{I},\textbf{v}}^{(0)}$ to $\mathcal{F}_{\mathcal{I},\textbf{v}}^{(j)}$ as
$$\phi((\textbf{f}_s^{(0)},\textbf{g}_0))=(\textbf{f}_{s'}^{(j)},\textbf{g}_j), \;\;\;\; (\textbf{f}_s^{(0)},\textbf{g}_0)\in\mathcal{F}_{\mathcal{I},\textbf{v}}^{(0)},\; (\textbf{f}_{s'}^{(j)},\textbf{g}_j)\in\mathcal{F}_{\mathcal{I},\textbf{v}}^{(j)},$$
where $f_{s,i}^{(0)}=f_{s',i}^{(j)}$ for $i\in[0,m-1]\setminus \{\xi_{0},\xi_{1},\cdots,\xi_{t-1}\}$ and $f_{s,{\xi_{h}}}^{(0)}+g_h^{(j)}(q-z)=f_{s',{\xi_{h}}}^{(j)}$ for $h\in [0,t-1]$. Given any $h\in [0,t-1]$, let us assume that $f_{s,{\xi_{h}}}^{(0)}=c_{h}+\eta_h$ for $\eta_h\in [1,q-z]$ and $f_{s',{\xi_{h}}}^{(j)}=c'_{h}+\eta'_h$. Based on {\it Construction 1}, we have $c_{h}=c'_{h}-g_h^{(j)}(q-z)$. This implies that $c'_{h}+\eta'_h=f_{s',{\xi_{h}}}^{(j)}=f_{s,{\xi_{h}}}^{(0)}+g_h^{(j)}(q-z)=c_{h}+\eta_h+g_h^{(j)}(q-z)=c'_{h}-g_h^{(j)}
(q-z)+\eta_h+g_h^{(j)}(q-z)=c'_{h}+\eta_h$, i.e., $\eta_h=\eta'_h$ holds.
Hence, given any $(\textbf{f}_{s}^{(0)},\textbf{g}_0)\in\mathcal{F}_{\mathcal{I},\textbf{v}}^{(0)}$, we can find a unique row $(\textbf{f}_{s'}^{(j)},\textbf{g}_j)\in\mathcal{F}_{\mathcal{I},\textbf{v}}^{(j)}$ such that $\phi((\textbf{f}_s^{(0)},\textbf{g}_0))=(\textbf{f}_{s'}^{(j)},\textbf{g}_j)$. This means that $\phi$ is indeed a mapping from $\mathcal{F}_{\mathcal{I},\textbf{v}}^{(0)}$ to $\mathcal{F}_{\mathcal{I},\textbf{v}}^{(j)}$. Further based on the definition of $\phi$, it can be seen that $\phi$ is an injective mapping from $\mathcal{F}_{\mathcal{I},\textbf{v}}^{(0)}$ to $\mathcal{F}_{\mathcal{I},\textbf{v}}^{(j)}$. Then we have $|\mathcal{F}_{\mathcal{I},\textbf{v}}^{(0)}|\leq|\mathcal{F}_{\mathcal{I},\textbf{v}}^{(j)}|$. Also, we can define a mapping $\phi'$ from $\mathcal{F}_{\mathcal{I},\textbf{v}}^{(j)}$ to $\mathcal{F}_{\mathcal{I},\textbf{v}}^{(0)}$ as
$$\phi'((\textbf{f}_{s'}^{(j)}),\textbf{g}_j)=(\textbf{f}_{s}^{(0)},\textbf{g}_0), \;\;\;
(\textbf{f}_{s'}^{(j)},\textbf{g}_j)\in\mathcal{F}_{\mathcal{I},\textbf{v}}^{(j)},
(\textbf{f}_s^{(0)},\textbf{g}_0)\in\mathcal{F}_{\mathcal{I},\textbf{v}}^{(0)},$$
%where $f_{s,i}^{(0)}=f_{s',i}^{(j)}$ for $i\in[0,m-1]\setminus \{\xi_{j-1}\}$ and $f_{s,{\xi_{j-1}}}^{(0)}+(q-z)=f_{s',{\xi_{j-1}}}^{(j)}$.
 where $f_{s',i}^{(j)}=f_{s,i}^{(0)}$ for $i\in[0,m-1]\setminus \{\xi_{0},\xi_{1},\cdots,\xi_{t-1}\}$ and $f_{s,{\xi_{h}}}^{(0)}+g_h^{(j)}(q-z)=f_{s',{\xi_{h}}}^{(j)}$ for $h\in [0,t-1]$.
 With the same analysis, we have $|\mathcal{F}_{\mathcal{I},\textbf{v}}^{(0)}|\geq|\mathcal{F}_{\mathcal{I},\textbf{v}}^{(j)}|$. Hence, we have $|\mathcal{F}_{\mathcal{I},\textbf{v}}^{(0)}|=|\mathcal{F}_{\mathcal{I},\textbf{v}}^{(j)}|$. For any $\mathcal{I}'\neq\mathcal{I}$, we also have $|\mathcal{F}_{\mathcal{I}',\textbf{v}}^{(0)}|=|\mathcal{F}_{\mathcal{I}',\textbf{v}}^{(j)}|$ with the similar argument above. Therefore, array $\textbf{P}_j$ satisfies the statement of (a).

%Let $\mathcal{D}=\{\textbf{g}_j+\textbf{x}_i\mid j\in[1, t], i\in[0,t-1]\}$, where $\textbf{x}_i=(0,\cdots,1,\cdots,0)$ is a vector with $i$th entry being 1 and others 0s. The computations are performed in modulo $\lfloor\frac{q-1}{q-z}\rfloor$. With the similar discussion above, based on $\textbf{P}_1,\textbf{P}_2,\cdots,\textbf{P}_t$, we can obtain other $\textbf{P}_{t+1}, \textbf{P}_{t+2}, \cdots \textbf{P}_{t+u}$ satisfying the statement of (a) with row indices generated from
%$\mathcal{F}^{(t+u')}_{\textbf{g}_{t+u'}}=\{(\textbf{f}_s^{(t+u')},\textbf{g}_{t+u'})\mid\textbf{f}_s^{(t+u')}\in \textbf{F}_{\textbf{g}_{t+u'}}, s\in[0, q^{m-1}-1]\},$
%where $\textbf{F}_{\textbf{g}_{t+u'}}=\textbf{F}_{\textbf{g}_{t+1}}$, $\sum_{r=0}^{m-1}f_{s,r}^{(t+1)}=2(q-z)$ for $s\in[0, q^{m-1}-1]$, $\textbf{g}_{t+u'}\in\mathcal{D}\backslash\{\textbf{g}_j\mid j\in[0,t]\}$,
% $u'\in[1,u]$ and $u$ is the cardinality of the set $\mathcal{D}\backslash\{\textbf{g}_j\mid j\in[0,t]\}$. Also, we can perform the above operations as many times as possible until we obtain any array $\textbf{P}_j$ in $\textbf{P}$ satisfying the statement of (a). This completes the proof of (a).

(b). According to the proof of (a), the statement of (b) is clear.
\end{proof}

Now let us prove {\it Theorem 1}.
\begin{proof}
Let \textbf{P} be the array generated from {\it Construction 1}. It can be observed that the vector \textbf{v} occurring in the column $(\mathcal{I},\textbf{c})$ of $\textbf{P}_j$ does not appear in the column $(\mathcal{I},\textbf{c})$ of $\textbf{P}_{j'}$ for $j'\neq j$ and $j, j'\in[0, \lfloor\frac{q-1}{q-z}\rfloor^{t}-1]$. Let us assume that the vector \textbf{v} occurring in the column $(\mathcal{I},\textbf{c})$ of $\textbf{P}_j$ appears in the column $(\mathcal{I},\textbf{c})$ of $\textbf{P}_{j'}$ for $j'\neq j$. Let $\textbf{g}_{j}=(g^{(j)}_0, g^{(j)}_1, \cdots, g^{(j)}_{t-1})$, $\textbf{g}_{j'}=(g^{(j')}_0, g^{(j')}_1, \cdots, g^{(j')}_{t-1})$ and $\textbf{c}=(c_0, c_1, \cdots, c_{t-1})$ be the vectors defined in {\it Construction 1}. Without loss of generality, we can assume that $g^{(j)}_i\neq g^{(j')}_i$ for some integer $i\in[0,t-1]$. Based on {\it Construction 1}, we have $c_i-g^{(j)}_i(q-z)=c_i-g^{(j')}_i(q-z)$. This is impossible since $g^{(j)}_i(q-z), g^{(j')}_i(q-z)\in[0,z-1]$. Based on {\it Proposition 3}, we also have that each vector \textbf{v} of \textbf{P} occurs in exactly ${m\choose t}\lfloor\frac{q-1}{q-z}\rfloor^{t}$ columns, and \textbf{v} occurs
the same number of times in each column that contains \textbf{v}, i.e., array \textbf{P} is a PDA such that the coding gain of each vector entry $(\textbf{v},o(\textbf{v}))$ is ${m\choose t}\lfloor\frac{q-1}{q-z}\rfloor^{t}$. Further based on {\it Proposition 2}, it can be seen that each column of \textbf{P} has $\lfloor\frac{q-1}{q-z}\rfloor^{t}[q^{m-1}-q^{m-t-1}(q-z)^{t}]$ $``\ast"$s and the number of non-star entries in $\textbf{P}$ is $\lfloor\frac{q-1}{q-z}\rfloor^{t}{m\choose t}q^{m-1}(q-z)^{t}$. This implies that the number of distinct vector entries in $\textbf{P}$ is
$\frac{\lfloor\frac{q-1}{q-z}\rfloor^{t}{m\choose t}q^{m-1}(q-z)^{t}}{{m\choose t}\lfloor\frac{q-1}{q-z}\rfloor^{t}}=q^{m-1}(q-z)^{t}$ since each vector entry $(\textbf{v},o(\textbf{v}))$ appears ${m\choose t}\lfloor\frac{q-1}{q-z}\rfloor^{t}$ times. Therefore, \textbf{P} is an $({m\choose t}q^{t}, \lfloor\frac{q-1}{q-z}\rfloor^{t}q^{m-1}, \lfloor\frac{q-1}{q-z}\rfloor^{t}[q^{m-1}-q^{m-t-1}(q-z)^{t}], q^{m-1}(q-z)^{t})$ PDA with a memory ratio of $\frac{M}{N}=1-(\frac{q-z}{q})^{t}$ and a transmission rate of $R=\frac{(q-z)^{t}}{\lfloor\frac{q-1}{q-z}\rfloor^{t}}$.
\end{proof}
\section{Proof of Theorem 2}
\begin{proof}Let $\hat{\textbf{P}}=(\hat{\textbf{P}}_{0}\;
     \cdots\;
     \hat{\textbf{P}}_{j}\;
     \cdots\;
     \hat{\textbf{P}}_{\lfloor\frac{q-1}{q-z}\rfloor^{t}-1})$ denote an array obtained from the transformation of {\it Construction 1}. It can be seen that it is impossible for a vector entry to appear more than once in each column of $\hat{\textbf{P}}$, since $\hat{\textbf{P}}_j$ is obtained by selecting some columns of $\textbf{P}_j'$ and $\textbf{P}_j'$ is a PDA (this can be checked with the similar proof of {\it Proposition 1}) for $j\in[0, \lfloor\frac{q-1}{q-z}\rfloor^{t}-1]$. Note that the first $m-1$ coordinates of the row indices of $\hat{\textbf{P}}_j$ are arranged in the lexicographic order from top to bottom and $\hat{\textbf{P}}_j$ $(j\neq0)$ is obtained by selecting the columns of $\textbf{P}_j'$ indexed by $(\mathcal{I}, \textbf{c})$, where $\mathcal{I}\in {[0,m-2]\choose t}$ and $\textbf{c}\in\{(c_0, c_1,\cdots, c_{t-1})\mid c_i\in [0,q-1], i\in[0, t-1]\}$.
Assume that vector entry $\textbf{e}=(\textbf{v},o(\textbf{v}))$ appears more than once in some row of $\hat{\textbf{P}}$. Without loss of generality, let us assume that $\hat{p}_{j (\textbf{f}_{s}^{(j)},\textbf{g}_j),(\mathcal{I},\textbf{c})}=\hat{p}_{j' (\textbf{f}_{s'}^{(j')},\textbf{g}_{j'}),(\mathcal{I'},\textbf{c}')}=\textbf{e}$ for $j\neq j'$, where
 $$(\textbf{f}_s^{(j)},\textbf{g}_j)=((f^{(j)}_{s,0},f^{(j)}_{s,1},\cdots,f^{(j)}_{s,m-2},f^{(j)}_{s,m-1}),(g^{(j)}_0, g^{(j)}_1, \cdots, g^{(j)}_{t-1}));$$
 $$(\textbf{f}_{s'}^{(j')},\textbf{g}_{j'})=((f^{(j)}_{s,0},f^{(j)}_{s,1},\cdots,f^{(j)}_{s,m-2},f^{(j')}_{s',m-1}),(g^{(j')}_0, g^{(j')}_1, \cdots, g^{(j')}_{t-1}));$$
 $$(\mathcal{I},\textbf{c})=(\{\xi_0,\xi_1,\cdots,\xi_{t-1}\},(c_0,c_1,\cdots,c_{t-1}));(\mathcal{I}',\textbf{c}')
 =(\{\xi'_0,\xi'_1,\cdots,\xi'_{t-1}\},(c'_0,c'_1,\cdots,c'_{t-1})).$$
%So in order to prove that there is no vector entry occurring more than once in each row of $\hat{\textbf{P}}$, we just need to
Let us consider the following two cases.

\textbf{Case 1:} $|\mathcal{I}\cap\mathcal{I}'|=t$ and $\mathcal{I}, \mathcal{I}'\in {[0,m-2]\choose t}$.
 % If $\textbf{F}_{\textbf{g}_j}\neq \textbf{F}_{\textbf{g}_{j'}}$, based on the definition of $\textbf{F}_{\textbf{g}_j}$, we have $a_{m-1}\neq a'_{m-1}$. This implies that $\hat{\textbf{P}}_{j (\textbf{f}_{s}^{(j)},\textbf{g}_j),(\mathcal{I},\textbf{c})}=\hat{\textbf{P}}_{j' (\textbf{f}_{s'}^{(j')},\textbf{g}_{j'}),(\mathcal{I'},\textbf{c}')}$ is impossible due to the entry rule of (15). Now let us consider  $\textbf{F}_{\textbf{g}_j}= \textbf{F}_{\textbf{g}_{j'}}$.
  Assume that vector entry \textbf{e} appears in row $(\textbf{f}_s^{(j)},\textbf{g}_j)=((f^{(j)}_{s,0},f^{(j)}_{s,1},\cdots,f^{(j)}_{s,m-2},f^{(j)}_{s,m-1}),(g^{(j)}_0, g^{(j)}_1, \cdots, g^{(j)}_{t-1}))$ of $\hat{\textbf{P}}_j$, then \textbf{e} must appear in row $(\textbf{f}_{s'}^{(j')},\textbf{g}_{j'})=((f^{(j')}_{s',0},f^{(j')}_{s',1},\cdots,f^{(j')}_{s',m-2},f^{(j')}_{s',m-1}),(g^{(j')}_0, g^{(j')}_1, \cdots, g^{(j')}_{t-1}))$ of $\hat{\textbf{P}}_{j'}$. We claim that $(f^{(j)}_{s,0},f^{(j)}_{s,1},\cdots,f^{(j)}_{s,m-2})\neq(f^{(j')}_{s',0},f^{(j')}_{s',1},\cdots,f^{(j')}_{s',m-2})$. This can be verified as follows. Suppose that \textbf{e} appears in row $((f^{(0)}_{s'',0},f^{(0)}_{s'',1},\cdots,f^{(0)}_{s'',m-2},f^{(0)}_{s'',m-1}),\textbf{g}_{0})$ of $\textbf{P}'_{0}$. According to $\textbf{g}_j\neq\textbf{g}_{j'}$, it can be seen that the hamming distance between the vectors $\textbf{g}_j$ and $\textbf{g}_{j'}$ is at least one. Without loss of generality, assume that there exists an integer $\alpha\in[0,t-1]$ such that $g^{(j)}_{\alpha}\neq g^{(j')}_{\alpha}$. Based on the process of the transform and the proof of {\it Proposition 3}, we obtain
$(f^{(j)}_{s,0},f^{(j)}_{s,1},\cdots,f^{(j)}_{s,m-2})=(f^{(0)}_{s'',0},\cdots,f^{(0)}_{s'',\xi_0}+g^{(j)}_0(q-z),\cdots,f^{(0)}_{s'',\xi_{\alpha}}
+g^{(j)}_{\alpha}(q-z),\cdots,f^{(0)}_{s'',\xi_{t-1}}+g^{(j)}_{t-1}(q-z),\cdots,f^{(0)}_{s'',m-2})$ and
$(f^{(j')}_{s,0},f^{(j')}_{s,1},\cdots,f^{(j')}_{s,m-2})=(f^{(0)}_{s'',0},\cdots,f^{(0)}_{s'',\xi_0}+g^{(j)}_0(q-z),\cdots,f^{(0)}_{s'',\xi_{\alpha}}
+g^{(j')}_{\alpha}(q-z),\cdots,f^{(0)}_{s'',\xi_{t-1}}+g^{(j)}_{t-1}(q-z),\cdots,f^{(0)}_{s'',m-2}).$
Then we have $f^{(0)}_{s'',\xi_{\alpha}}+g^{(j)}_{\alpha}(q-z)\neq f^{(0)}_{s'',\xi_{\alpha}}+g^{(j')}_{\alpha}(q-z)$ since $g^{(j)}_{\alpha}(q-z), g^{(j')}_{\alpha}(q-z)\in[0,z-1]$. This implies $(f^{(j)}_{s,0},f^{(j)}_{s,1},\cdots,f^{(j)}_{s,m-2})\neq(f^{(j')}_{s',0},f^{(j')}_{s',1},\cdots,f^{(j')}_{s',m-2})$.

%This can be verified as follows. If $\textbf{g}_j,\textbf{g}_{j'}\in\mathcal{E}_p$ and $\textbf{P}'_j$,$\textbf{P}'_{j'}$ are both generated from $\textbf{P}'_{i}$, where $p\in[1,\lfloor\frac{q-1}{q-z}\rfloor^{t}-1]$ and $j,j'>i$, without loss of generality, assume that \textbf{e} occurs in row $((a_0,a_1,\cdots,a_{m-2},a_{m-1}),\textbf{g}_i)$ of $\textbf{P}'_{i}$. Then \textbf{e} must occur in row $((a_0,\cdots,a_u+(q-z),\cdots,a_{m-2},a_{m-1}),\textbf{g}_j)$ of $\textbf{P}'_{j}$ and row $((a_0,\cdots,a_{u'}+(q-z),\cdots,a_{m-2},a_{m-1}),\textbf{g}_{j'})$ of $\textbf{P}'_{j'}$ since $\textbf{g}_j\neq \textbf{g}_{j'}$ and $\textbf{g}_j=\textbf{g}_i+\textbf{x}_u$, $\textbf{g}_{j'}=\textbf{g}_i+\textbf{x}_{u'}$. If $\textbf{g}_j,\textbf{g}_{j'}\in\mathcal{E}_p$ and $\textbf{P}'_j$,$\textbf{P}'_{j'}$ are generated from $\textbf{P}'_{i}$ and $\textbf{P}'_{i'}$, respectively, where $p\in[1,\lfloor\frac{q-1}{q-z}\rfloor^{t}-1]$ and $j>i,j'>i'$, without loss of generality, assume that \textbf{e} occurs in row $((a_0,a_1,\cdots,a_{m-2},a_{m-1}),\textbf{g}_i)$ of $\textbf{P}'_{i}$ and row $((b_0,b_1,\cdots,b_{m-2},b_{m-1}),\textbf{g}_{i'})$ of $\textbf{P}'_{i'}$. Then the hamming distance between the vectors $(a_0,a_1,\cdots,a_{m-2},a_{m-1})$ and $(b_0,b_1,\cdots,b_{m-2},b_{m-1})$ is at least two. Without loss of generality, suppose $a_0\neq b_0$ and $a_1\neq b_1$

\textbf{Case 2:} $|\mathcal{I}\cap\mathcal{I}'|<t$ and $\mathcal{I}'\in {[0,m-2]\choose t}$.
%If $|\mathcal{I}\cap\mathcal{I}'|<t$ and $\mathcal{I}'\subseteq {[0,m-2]\choose t}$, this implies that
There must exist two distinct integers, say $\alpha$, $\alpha'\in[0, m-1]$, satisfying $\alpha\in\mathcal{I}, \alpha\notin\mathcal{I}'$ and $\alpha'\in\mathcal{I}', \alpha'\notin\mathcal{I}$. Without loss of generality, let us assume that $\alpha=\xi_0$. Based on the construction of $\hat{\textbf{P}}_j$ and $\hat{\textbf{P}}_{j'}$,
we obtain $f^{(j)}_{s,\xi_0}=c_0-g^{(j)}_0(q-z)$. This implies $\hat{p}_{j (\textbf{f}_{s}^{(j)},\textbf{g}_j),(\mathcal{I},\textbf{c})}=\ast$
 since $f^{(j)}_{s,\xi_0}=c_0-g^{(j)}_0(q-z)\in\{c_0, c_0-1, \cdots, c_0-(z-1)\}$, which contradicts the hypothesis.

Based on the above argument, it can be seen that C3-(a) of {\it Definition 1} holds. It remains to show that C3-(b) of {\it Definition 1} holds for $\hat{\textbf{P}}$. Note that $\hat{\textbf{P}}_{j}$ is a PDA for $j\in[1,\lfloor\frac{q-1}{q-z}\rfloor^{t}-1]$.
%So in order to prove that C3-(b) of Definition 1 holds for $\hat{\textbf{P}}$,
To do this we just need to prove if $\hat{p}_{j (\textbf{f}_{s}^{(j)},\textbf{g}_j),(\mathcal{I},\textbf{c})}=\hat{p}_{j' (\textbf{f}_{s'}^{(j')},\textbf{g}_{j'}),(\mathcal{I'},\textbf{c}')}=\textbf{e}$ for $j\neq j'$ and $\sum_{i=0}^{t-1}g^{(j)}_i\geq\sum_{i=0}^{t-1}g^{(j')}_i$, then $\hat{p}_{j (\hat{\textbf{f}}_{s'}^{(j')},\textbf{g}_{j}),(\mathcal{I},\textbf{c})}=\hat{p}_{j' (\hat{\textbf{f}}_{s}^{(j)},\textbf{g}_{j'}),(\mathcal{I'},\textbf{c}')}=\ast$, where
 $$(\textbf{f}_s^{(j)},\textbf{g}_j)=((f^{(j)}_{s,0},f^{(j)}_{s,1},\cdots,f^{(j)}_{s,m-2},f^{(j)}_{s,m-1}),(g^{(j)}_0, g^{(j)}_1, \cdots, g^{(j)}_{t-1}));$$
 $$(\textbf{f}_{s'}^{(j')},\textbf{g}_{j'})=((f^{(j')}_{s',0},f^{(j')}_{s',1},\cdots,f^{(j')}_{s',m-2},f^{(j')}_{s',m-1}),(g^{(j')}_0, g^{(j')}_1, \cdots, g^{(j')}_{t-1}));$$
 $$(\mathcal{I},\textbf{c})=(\{\xi_0,\xi_1,\cdots,\xi_{t-1}\},(c_0,c_1,\cdots,c_{t-1}));(\mathcal{I}',\textbf{c}')
 =(\{\xi'_0,\xi'_1,\cdots,\xi'_{t-1}\},(c'_0,c'_1,\cdots,c'_{t-1}));$$
 $$\hat{\textbf{f}}_{s'}^{(j')}=(f^{(j')}_{s',0},f^{(j')}_{s',1},\cdots,f^{(j')}_{s',m-2},f^{(j')}_{s',m-1}+x(q-z));
 \hat{\textbf{f}}_{s}^{(j)}=(f^{(j)}_{s,0},f^{(j)}_{s,1},\cdots,f^{(j)}_{s,m-2},f^{(j)}_{s,m-1}-x(q-z));$$
\[x=\sum_{i=0}^{t-1}g^{(j)}_i-\sum_{i=0}^{t-1}g^{(j')}_i.\]
Let us consider the following two cases.

\textbf{Case 3:} $\mathcal{I}, \mathcal{I}'\in {[0,m-2]\choose t}$. Since $(
     \hat{\textbf{P}}_{j};
     \hat{\textbf{P}}_{j'})$
 is a PDA, we have $\hat{p}_{j' (\textbf{f}_{s'}^{(j')},\textbf{g}_{j'}),(\mathcal{I},\textbf{c})}=\hat{p}_{j (\textbf{f}_{s}^{(j)},\textbf{g}_{j}),(\mathcal{I'},\textbf{c}')}=\ast$. This implies $\hat{p}_{j (\hat{\textbf{f}}_{s'}^{(j')},\textbf{g}_{j}),(\mathcal{I},\textbf{c})}=\hat{p}_{j' (\hat{\textbf{f}}_{s}^{(j)},\textbf{g}_{j'}),(\mathcal{I'},\textbf{c}')}=\ast$ due to the entry rule in {\it Construction 1}.

\textbf{Case 4:} $\mathcal{I}\in {[0,m-2]\choose t}$ and $\mathcal{I}'\not\in {[0,m-2]\choose t}$. In this case, we just need to consider $j'=0$, i.e., we just need to prove if
 $\hat{p}_{j (\textbf{f}_{s}^{(j)},\textbf{g}_j),(\mathcal{I},\textbf{c})}=\hat{p}_{0 (\textbf{f}_{s'}^{(0)},\textbf{g}_{0}),(\mathcal{I'},\textbf{c}')}=\textbf{e}$ for $j\neq 0$, then $\hat{p}_{j (\hat{\textbf{f}}_{s'}^{(0)},\textbf{g}_{j}),(\mathcal{I},\textbf{c})}=\hat{p}_{0 (\hat{\textbf{f}}_{s}^{(j)},\textbf{g}_{0}),(\mathcal{I'},\textbf{c}')}=\ast$, where
  $\textbf{g}_0=(g^{(0)}_0, g^{(0)}_1, \cdots, g^{(0)}_{t-1})=(0,0,\cdots,0)$.
  With the similar argument of Case 3, it can be seen that $\hat{p}_{j (\hat{\textbf{f}}_{s'}^{(0)},\textbf{g}_{j}),(\mathcal{I},\textbf{c})}=\ast$
   holds. Next we show that $\hat{p}_{0 (\hat{\textbf{f}}_{s}^{(j)},\textbf{g}_{0}),(\mathcal{I}',\textbf{c}')}=\ast$
   also holds. If $|\mathcal{I}\cap\mathcal{I}'|<t-1$,
 based on entry rule of $\hat{\textbf{P}}$ and {\it Construction 1}, there must exist an integer $\xi_{\alpha}\in\mathcal{I}'$ such that $f^{(0)}_{s',\xi_{\alpha}}=c'_{\alpha}-g^{(0)}_{\alpha}(q-z)$. This implies $\hat{p}_{0 (\hat{\textbf{f}}_{s}^{(j)},\textbf{g}_{0}),(\mathcal{I}',\textbf{c}')}=\ast$ since $f^{(0)}_{s',\xi_{\alpha}}=c'_{\alpha}$.
 If $|\mathcal{I}\cap\mathcal{I}'|=t-1$, without loss of generality, let us assume that $\xi_{\alpha_0}=\xi'_{\beta_0}, \xi_{\alpha_1}=\xi'_{\beta_1},\cdots,\xi_{\alpha_{t-2}}=\xi'_{\beta_{t-2}}$, where $\{\xi_{\alpha_0},\xi_{\alpha_1},\cdots,\xi_{\alpha_{t-2}}\}\subseteq\mathcal{I}$ and $\{\xi'_{\beta_0},\xi'_{\beta_1},\cdots,\xi'_{\beta_{t-2}}\}\subseteq\mathcal{I}'$.
 Based on {\it Construction 1}, we have
  \begin{align*}
 &c_{\alpha_0}-g^{(j)}_{\alpha_0}(q-z)=c'_{\beta_0}-g^{(0)}_{\beta_0}(q-z)=c'_{\beta_0};\\
 &c_{\alpha_1}-g^{(j)}_{\alpha_1}(q-z)=c'_{\beta_1}-g^{(0)}_{\beta_1}(q-z)=c'_{\beta_1};\\
 &\vdots\\
 &c_{\alpha_{t-2}}-g^{(j)}_{\alpha_{t-2}}(q-z)=c'_{\beta_{t-2}}-g^{(0)}_{\beta_{t-2}}(q-z)=c'_{\beta_{t-2}};\\
 &f^{(j)}_{s,m-1}=c'_{t-1}-g^{(0)}_{t-1}(q-z)=c'_{t-1}.
 \end{align*}
Suppose that $f^{(j)}_{s,\xi'_{\beta_i}}=c'_{\beta_i}+\gamma$ for $\gamma\in[1,q-z]$ and $i\in[0,t-2]$, i.e., $\hat{p}_{0 (\hat{\textbf{f}}_{s}^{(j)},\textbf{g}_{0}),(\mathcal{I}',\textbf{c}')}\neq\ast$. Then we obtain
 \begin{align*}
  &c_{\alpha_0}-g^{(j)}_{\alpha_0}(q-z)=c'_{\beta_0}=f^{(j)}_{s,\xi'_{\beta_0}}-\gamma;\\
% c_{\alpha_1}-g^{(j)}_{\alpha_1}(q-z)=c'_{\beta_1}=f^{(j)}_{s,\xi'_{\beta_1}}-\gamma;\\
 &\vdots\\
 &c_{\alpha_{t-3}}-g^{(j)}_{\alpha_{t-3}}(q-z)=f^{(j)}_{s,\xi'_{\beta_{t-3}}}-\gamma;\\
& c_{\alpha_{t-2}}-g^{(j)}_{\alpha_{t-2}}(q-z)=f^{(j)}_{s,\xi'_{\beta_{t-2}}}-\gamma.
 \end{align*}
  This implies
 $c_{\alpha_i}+\gamma-g^{(j)}_{\alpha_i}(q-z)=f^{(j)}_{s,\xi'_{\beta_i}}=f^{(j)}_{s,\xi_{\alpha_i}}$ for $i\in[0,t-2]$. If there exists an integer $i\in$\\$[0,t-2]$ such that $g^{(j)}_{\alpha_i}\neq0$, then $\hat{p}_{j (\textbf{f}_{s}^{(j)},\textbf{g}_j),(\mathcal{I},\textbf{c})}=\ast$ due to $g^{(j)}_{\alpha_i}(q-z)\in[q-z,z-1]$ and $q<2z$, which contradicts $\hat{p}_{j (\textbf{f}_{s}^{(j)},\textbf{g}_j),(\mathcal{I},\textbf{c})}=\textbf{e}$. So $\gamma\notin [1,q-z]$, i.e., $\hat{p}_{0 (\hat{\textbf{f}}_{s}^{(j)},\textbf{g}_{0}),(\mathcal{I}',\textbf{c}')}=\ast$. If $g^{(j)}_{\alpha_i}=0$ for any $i\in[0,t-2]$, it can be seen that
 $g^{(j)}_{t-1}\neq0$ since $\textbf{g}_j\neq \textbf{g}_0=(0,0,\cdots,0)$. Note that
 $f^{(j)}_{s,m-1}=c'_{t-1}$.
 This implies $f^{(j)}_{s,m-1}-x(q-z)\in\{c'_{t-1},c'_{t-1}-1,\cdots,c'_{t-1}-(z-1)\}$, where $x=\sum_{i=0}^{t-1}g^{(j)}_i-\sum_{i=0}^{t-1}g^{(0)}_i=g^{(j)}_{t-1}$. Hence, we also have $\hat{p}_{0 (\hat{\textbf{f}}_{s}^{(j)},\textbf{g}_{0}),(\mathcal{I}',\textbf{c}')}=\ast$. Therefore, C3-(b) of {\it Definition
1} holds.

Since $\hat{\textbf{P}}=(
     \hat{\textbf{P}}_{0}\;
     %\hat{\textbf{P}}_{1}\;
     \cdots\;
     \hat{\textbf{P}}_{j}\;
     \cdots\;
     \hat{\textbf{P}}_{\lfloor\frac{q-1}{q-z}\rfloor^{t}-1})$ and $\hat{\textbf{P}}_{j}$ is a PDA for $j\in[0,\lfloor\frac{q-1}{q-z}\rfloor^{t}-1]$, it can be seen that the number of $``\ast"$s in each column of $\hat{\textbf{P}}$ is $q^{m-1}-q^{m-t-1}(q-z)^{t}$. Furthermore, based on {\it Proposition 3}, it can be seen that the number distinct entries in $\hat{\textbf{P}}$ is $q^{m-1}(q-z)^{t}$. Therefore, array $\hat{\textbf{P}}$ is an $(({m-1\choose t}\lfloor\frac{q-1}{q-z}\rfloor^{t}+{m\choose t}-{m-1\choose t})q^{t}$, $q^{m-1}$, $q^{m-1}-q^{m-t-1}(q-z)^{t}$, $q^{m-1}(q-z)^{t})$ PDA with a memory ratio of $\frac{M}{N}=1-(\frac{q-z}{q})^{t}$ and a transmission rate of $R=(q-z)^{t}$.
\end{proof}

%%%%%%%%%%%%%%%%%%%%%%
\bibliographystyle{ieeetr}

\begin{thebibliography}{}

\bibitem{1} M. A. Maddah-Ali and U. Niesen, ``Fundamental limits of caching,'' {\em IEEE Trans. Inf. Theory}, vol. 60, no. 5, pp. 2856-2867, May 2014.
 \bibitem{2}K. Wan, D. Tuninetti, and P. Piantanida, ``An index coding approach to caching with uncoded cache placement,'' {\em IEEE Trans. Inf. Theory }, vol. 66, no. 3, pp. 1318-1332, Mar. 2020.
\bibitem{3} H. Ghasemi and A. Ramamoorthy, ``Improved lower bounds for coded caching," {\em IEEE Trans. Inf. Theory}, vol. 63, no. 7, pp. 4388-4413, Jul. 2017.
%Q. Yu, M. A. Maddah-Ali, and A. S. Avestimehr, ``Characterizing the rate-memory tradeoff in cache networks within a factor of 2,'' {\em IEEE Trans. Inf. Theory }, vol. 65, no. 1, pp. 647-663, Jan. 2019.
\bibitem{20} S. Agrawal, K. V. Sushena Sree, and P. Krishnan,``Coded caching based on combinatorial designs," in Proc.
{\em IEEE Int. Symp. Inf. Theory (ISIT)}, Paris, France, Jul. 2019, pp. 1227-1231.
     \bibitem{22}  H. H. S. Chittoor, M. Bhavana, and P. Krishnan, ``Coded caching via projective geometry: A new low subpacketization scheme," in Proc. {\em IEEE Int. Symp. Inf. Theory (ISIT)}, Paris, France, Jul. 2019, pp. 682-686.
\bibitem{21} H. H. S. Chittoor, P. Krishnan, K. V. S. Sree, and B. Mamillapalli, ``Subexponential and linear subpacketization coded caching via projective geometry,'' {\em IEEE Trans. Inf. Theory}, vol. 67, no. 9, pp. 6193-6222, Sep. 2021.
\bibitem{16}  C. Shangguan, Y. Zhang, and G. Ge, ``Centralized coded caching schemes: A hypergraph theoretical approach," {\em IEEE Trans. Inf. Theory}, vol. 64, no. 8, pp. 5755-5766, Aug. 2018.
\bibitem{17}  K. Shanmugam, A. M. Tulino, and A. G. Dimakis, ``Coded caching with linear subpacketization is possible using Ruzsa-Szem\'{e}redi graphs," in Proc. {\em IEEE Int. Symp. Inf. Theory (ISIT)}, Aachen, Germany, Jun. 2017, pp. 1237-1241.
 \bibitem{18}  K. Shanmugam, A. G. Dimakis, J. Llorca, and A. M. Tulino, ``A unified Ruzsa-Szemer\'{e}di framework for finite-length coded caching," in Proc. The 51st ACSSC, Pacific Grove, CA, Oct. 2017, pp. 631-635.
 \bibitem{26}  K. Shanmugam, M. Ji, A. M. Tulino, J. Llorca, and A. G. Dimakis, ``Finite length analysis of caching-aided coded multicasting,"{\em IEEE Trans. Inf. Theory}, vol. 62, no. 10, pp. 5524-5537, Oct. 2016.
 \bibitem{13} Q. Yan, M. Cheng, X. Tang, and Q. Chen, ``On the placement delivery
array design for centralized coded caching scheme,'' {\em IEEE Trans. Inf. Theory}, vol. 63, no. 9, pp. 5821-5833, Sep. 2017.
\bibitem{14}  Q. Yan, X. Tang, Q. Chen, and M. Cheng, ``Placement delivery array design through strong edge coloring of bipartite graphs," {\em IEEE Commun. Lett.}, vol.22, no. 2, pp. 236-239, Feb. 2018.

    \bibitem{50} L. Tang and A. Ramamoorthy, ``Coded caching schemes with reduced
subpacketization from linear block codes," {\em IEEE Trans. Inf. Theory}, vol. 64, no. 4, pp. 3099-3120, Apr. 2018.
\bibitem{39}
M. Cheng, J. Jiang, X. Tang, and Q. Yan, ``Some variant of known coded caching schemes with good performance," {\em IEEE Trans. Commun.}, vol. 68, no.3, pp. 1370-1377, Mar., 2020.
\bibitem{19}  M. Cheng, J. Jiang, Q. Yan, and X. Tang, ``Constructions of coded caching schemes with flexible memory size," {\em IEEE Trans. Commun.}, vol. 67, no. 6, pp. 4166-4176, Jun. 2019.
    \bibitem{29} M. Cheng, J. Wang, X. Zhong, and Q. Wang, ``A framework of constructing placement delivery arrays for centralized coded caching," {\em IEEE Trans. Inf. Theory}, vol. 67, no. 11, pp. 7121-7131, Nov. 2021.
         \bibitem{24} M. Zhang, M. Cheng, J. Wang, and X. Zhong, ``Improving placement delivery array coded caching schemes with coded placement," {\em IEEE Access}, vol. 8, pp. 217456-217462, Dec. 2020.
   \bibitem{23} M. Cheng ,J. Jiang, Q. Wang, and Y. Yao, ``A generalized grouping scheme in
coded caching," {\em IEEE Trans. Commun.}, vol. 67, no. 5, pp. 3422-3430, May 2019.
%\bibitem{23}  M. Cheng, W. Zhang, and J. Jiang, ``So me new coded caching schemes with smaller subpacketization via Some known results," in Proc. IEEE ACCESS, vol. 8, no. 1, pp. 86305-86315, 21 May 2020.
\bibitem{31}   X. Zhong, M. Cheng, and J. Jiang, ``Placement delivery array based on
concatenating construction," {\em IEEE Commun. Lett.}, vol. 24, no. 6, pp. 1216-1220, Jun. 2020.
%\bibitem{32} W. Song, K. Cui, and L. Shi, (Oct. 2019), ``Some new constructions of coded caching schemes with reduced subpacketization," [Online]. Available: https://arXiv:1908.06570v2.
\bibitem{33} J. Michel and Q. Wang, ``Placement delivery arrays from combinations
of strong edge colorings," {\em IEEE Trans. Commun.}, vol. 68, no. 10, pp. 5953-5964, Oct. 2020.
       \bibitem{30} D. R. Stinson, {\em Combinatorial Designs: Construction and Analysis}, Springer, 2003, New York.



\end{thebibliography}

%\end{sloppypar}
%\end{spacing}
\end{document}